\documentclass{aa}  

\usepackage{graphicx}
\usepackage{txfonts}
\usepackage{lipsum}
\usepackage{subcaption}         
\usepackage{lscape}             
\usepackage{placeins}           
\usepackage{multirow}                

\begin{document}

   \title{Diagnosing the solar atmosphere through the Mg I b$_2$ 5173 \AA\ line}

   \subtitle{II. Morphological classification of the intensity and circular polarization profiles}

   \author{A. L. Siu-Tapia \inst{1,2}
        \and L. R. Bellot Rubio \inst{1,2}
        \and D. Orozco Su\'arez \inst{1,2}
        \and R. Gafeira \inst{3,4}
        }

   \institute{Instituto de Astrof\'isica de Andaluc\'ia (IAA-CSIC), Apdo. 3004, 18080 Granada, Spain\\
             \email{azaymisiu@gmail.com}
            \and Spanish Space Solar Physics Consortium, Spain\\
            \and Geophysical and Astronomical Observatory, Faculty of Science and Technology, University of Coimbra, Portugal\\
            \and Instituto de Astrof\'isica e Ci\^encias do Espa\c{c}o, Department of Physics, University of Coimbra, Portugal}

   \date{Received November 29, 2024}

 
  \abstract
   {The Mg I b$_2$ line at 5173 \AA\ is primarily magnetically sensitive to heights between the mid photosphere
and the low chromosphere, a region that has not been sufficiently explored in the solar atmosphere
but is crucial for understanding the magnetic coupling between the two layers. New generation solar
observatories, both space-borne and ground-based, are now performing polarimetric observations of this
spectral line, enabling simultaneous measurements with multiple spectral lines. This allows for
detailed studies of the magnetism around the temperature minimum region of the solar atmosphere at
high spatial, temporal, and spectral resolutions. }
   {We present a first classification of the Stokes
$I$ and $V$ profiles of the Mg I b$_2$ line using high spatial resolution observations from the Swedish 1-m
Solar Telescope.} 
   {We used the Euclidean distance to perform a morphological classification of the intensity and circular polarization profiles of the Mg I b$_2$ line. The physical properties of the resulting classes were analyzed using classical inference methods. Additionally, we present a two-line full-Stokes inversion of the representative profiles in which the Mg I b$_2$ line is treated fully under non-local thermodynamic equilibrium (NLTE) conditions, while the Fe I 6173 \AA\ line is simultaneously inverted under LTE assumptions to provide photospheric constraints. This approach offers insights into the temperature stratification and other physical gradients involved in the formation of the different profile morphologies.}
   {
   The intensity and polarization profiles of the Mg I b$_2$
line show clear spectral signatures depending on the magnetic environment and the variety of physical
processes occurring in the different solar magnetic regimes within the line formation range. We found nine classes of Stokes $V$ profiles and 
16 classes of Stokes $I$ profiles in our Mg I b$_2$ dataset. These classes can be further grouped into families based on shared characteristics,  physical properties, and location.}
   {Our classification provides important information on the different environments and processes occurring in the solar atmosphere around the temperature minimum region. It is also relevant for
improving the performance of NLTE inversions.}


   \maketitle

\section{Introduction}
\nolinenumbers

The Mg I b$_2$ line near 5173 \AA\ is a magnetically sensitive line whose core is formed under conditions of non-local thermodynamic equilibrium (NLTE),
mainly between 600 and 700 km above the solar surface \citep[e.g.,][]{Dorantes2022}. However, this line covers a large spectral width and thus has contributions from different layers of the solar atmosphere. In particular, the wings of the line have a primarily photospheric origin, while the core of the line can probe heights up to 900 km above the solar surface \citep{Mauas1988}. The latter has been shown to be mainly sensitive around the temperature minimum height in magnetic elements \citep{Briand1998}.
Therefore, despite its relatively weak polarization signals, especially in linear polarization \citep[e.g.,][]{delaCruz2010}, the Mg I b$_2$ line offers an opportunity to probe the magnetism of the solar atmosphere in a key region that includes the transition between the photosphere and the chromosphere.

The new generation of 4-m solar telescopes, such as the Daniel K. Inouye Solar Telescope \citep[DKIST;][]{Rimmele2019} and the European Solar Telescope \citep[EST;][]{Quintero2022} as well as the third flight of the SUNRISE solar balloon-borne observatory 
\citep{Lagg2024}, will provide spectropolarimetric measurements in a wide range of the solar spectrum, including the Mg I b$_2$ line in the visible, with greatly improved resolutions. This will allow the magnetic field and the plasma properties to be studied  simultaneously in different layers of the solar atmosphere at scales never observed before.
On the eve of these unprecedented observations, it is of great importance to gain a better understanding of the main characteristics and capabilities of the Mg I b$_2$ line. To date, 
its specific spectral and polarization signatures  have not yet been characterized. 

In this work, we perform the first survey of the intensity and circular polarization profiles of the Mg I b$_2$ 5173 \AA\ line over a bipolar magnetic region and its surrounding quiet Sun using high spatial resolution observations. We focus on the classification of the main intensity and polarization signatures  with the aim of identifying the full variety of Stokes $I$ and $V$ profile shapes emanating from the observed region. We also analyze the spatial distribution of the signals over specific structures in order to gain a better understanding of the physical processes that give rise to them. 
The characterization of the observed profiles
provides a wealth of information, including aspects about temperature and gradients along the line-of-sight (LOS) in the magnetic field, velocity, and density in the line-forming region, which can be used to improve the performance and speed of NLTE Stokes inversions.

\section{Observations} \label{sec:style}

We used spectropolarimetric observations of the Mg I b$_2$ line at 5173 \AA\ from the CRisp Imaging SpectroPolarimeter \citep[CRISP;][]{Scharmer2008} at the Swedish 1-meter Solar Telescope \citep[SST;][]{Scharmer2003}. The spectral line was sampled in a total of 14 wavelength points from $-500$ m\AA\ to $+500$ m\AA\ around the central wavelength in steps of 100 m{\AA}, except within the line core from $-100$ m\AA\ to $+100$ m{\AA}, where steps of 50 m\AA\ were employed, and an additional point closer to the continuum at $-700$ m{\AA}. 

As described in \citet[][hereafter Paper I]{Siu_paperI}, 
the dataset consists of four multi-wavelength temporal scans with a cadence of nearly 116 seconds between consecutive maps, which cover a field-of-view (FOV) of  $\sim 55\arcsec\times60\arcsec$ at a heliocentric angle of $\sim 26^{\circ}$ and with a spatial resolution of $0.057\arcsec$ per pixel. Four spectral lines were observed, two in polarimetric mode (Fe I 6173 \AA\ and Mg I b$_2$ 5173 {\AA}) and two in non-polarimetric mode (Ca II 8542 \AA\ and H$\alpha$ 6563 {\AA}). 
Further details can be found in Table 1 of Paper I.

 The data were reduced using the SSTRED pipeline \citep{delaCruz2015}, and the multi-object, multi-frame blind-deconvolution \citep[MOMFBD;][]{Vannoort2005} technique was used for image restoration.
 Subsequently, the data were subject to the usual post-processing steps: residual fringes and polarimetric cross-talk correction, and spectral and spatial gradient correction. The wavelength calibration was done by using the central wavelength of the average intensity profile in the entire dataset as a reference. In this work, we focus only on the morphological analysis of the Mg I b$_2$ line, but we use the full Stokes information of the Fe I 6173 \AA\ line as photospheric context for performing NLTE inversions.
 
 Finally, the data were convolved with a $3\times3$ low-pass filter kernel, resulting in a reduced polarization noise level of the order of $10^{-3}$. 
Nonetheless, most of the observed linear polarization signals, both in Stokes $Q$ and $U$, remain of the same order of magnitude as $\sigma_{Q,U}$. Therefore, the morphological classification in this work is limited to the Stokes $I$ and $V$ profiles  of the Mg I b$_2$ line.

 \begin{figure*}[ht!]
   \centering

  \includegraphics[width=\hsize]{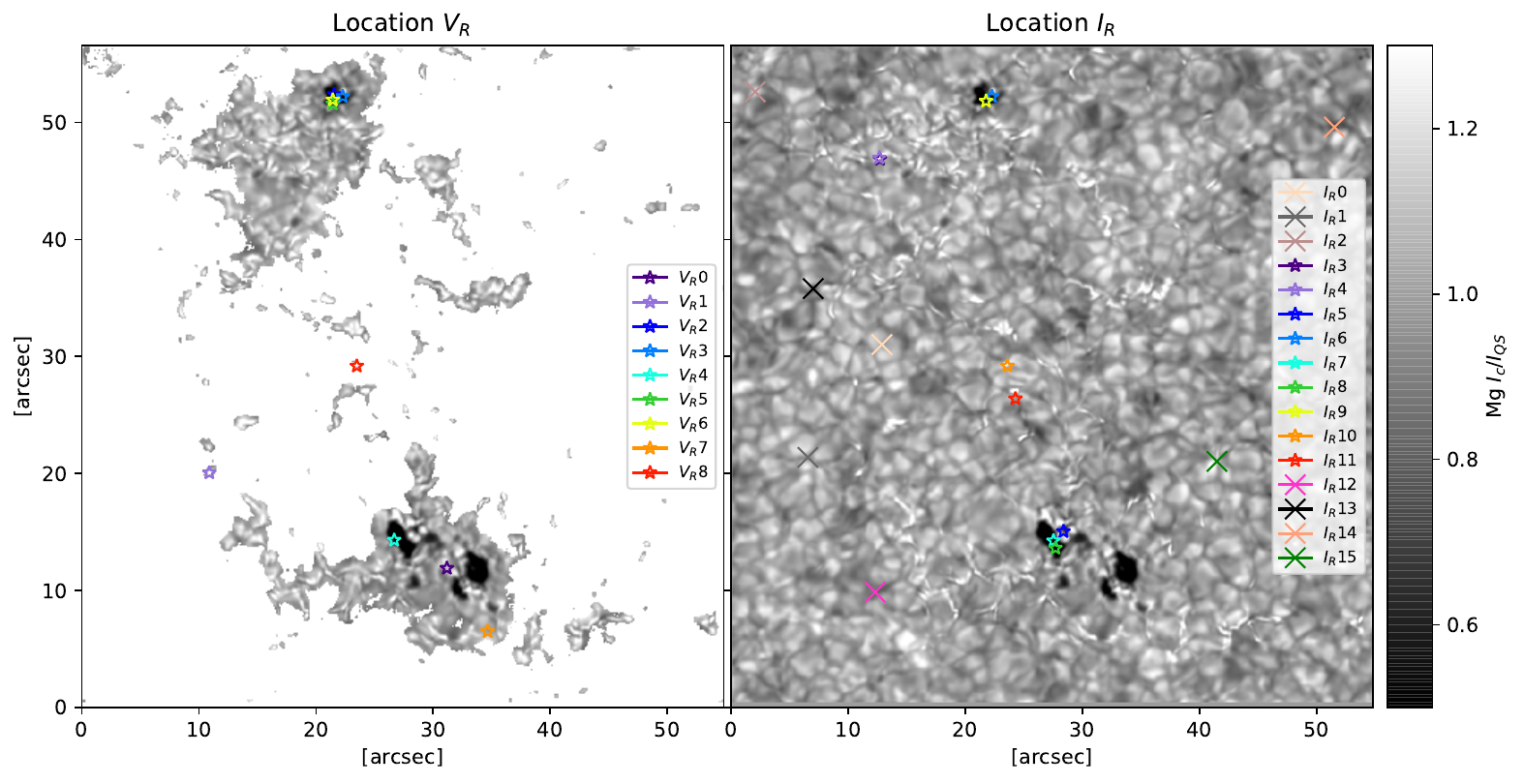}      
      \caption{Location of the nine representative profiles for Stokes $V$ (left) and of the 16 representative profiles for Stokes $I$ (right) on the  Mg intensity maps normalized to the mean quiet Sun value $I_{QS}$ (at -700 m{\AA}). The left panel shows only regions where the circular polarization signals are above the 5$\sigma$ level.
                }
         \label{fig:4bef}
   \end{figure*}

\begin{figure*}[ht!]
   \centering
      
      \includegraphics[width=\hsize]{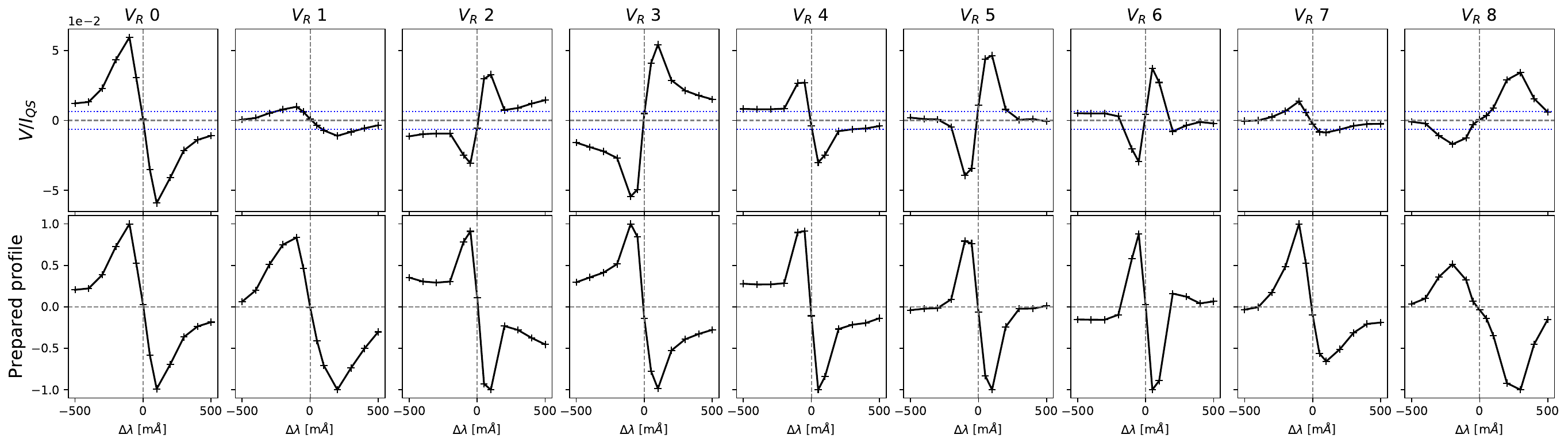}
      \caption{Nine Stokes $V$ representative profiles from the Mg dataset. 
The top panels show the profiles normalized to  
 $I_{QS}$ (at -700 m{\AA}). The gray dashed lines were placed at zero and the blue dotted lines show the $\pm5\sigma$ level.  
The bottom panels show the representative profiles after Doppler-shift removal, renormalization, and polarity adjustment as explained in the main text.    }
         \label{fig:4}
   \end{figure*}

 \begin{figure*}[ht!]
   \centering
      
          \includegraphics[width=1\hsize]{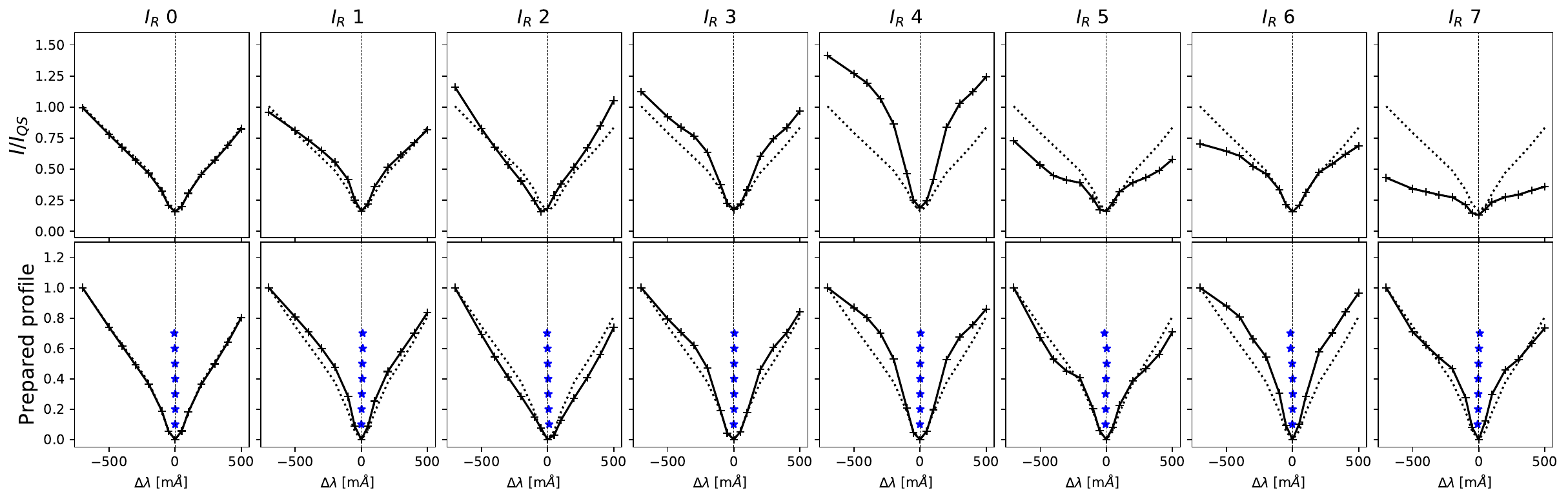}
          \includegraphics[width=1\hsize]{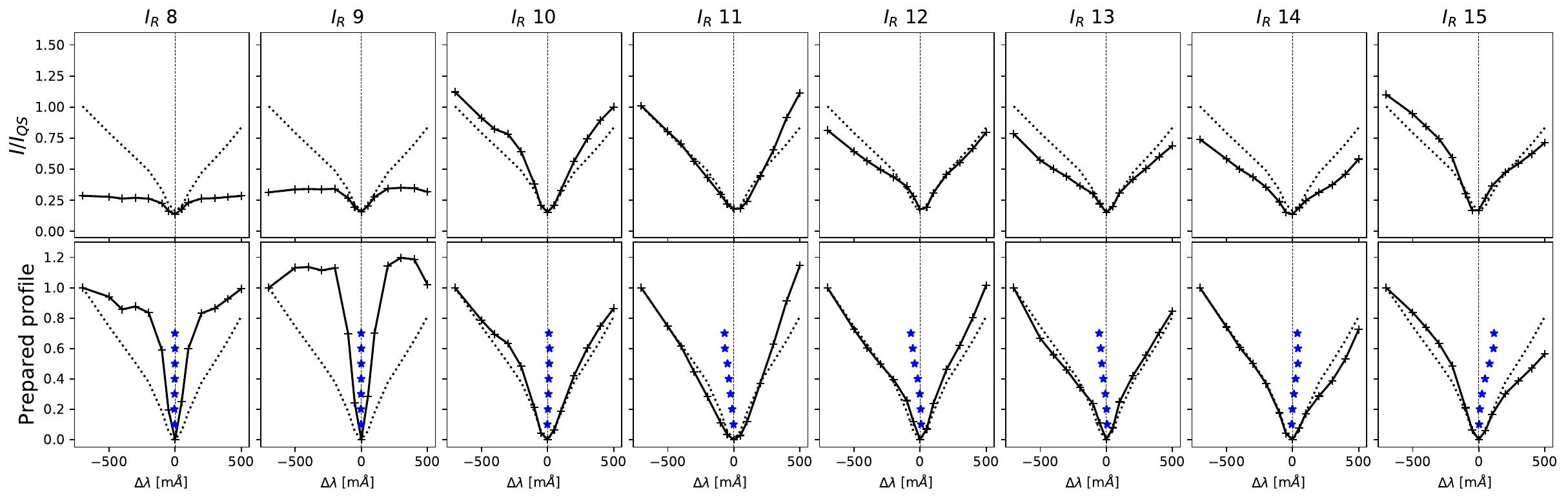}
      \caption{
      Sixteen Stokes $I$ representatives from the Mg dataset.
Same format as in Fig.\ \ref{fig:4}.
 The dotted lines show the average profile of the quiet Sun. The blue symbols in the bottom panels show the line bisectors 
 from 10$\%$ to 70$\%$ of the normalized continuum intensity.              }
         \label{fig:5}
   \end{figure*}


\section{Methods}

\subsection{Preparation of the dataset}
In order to perform the classification, the profiles were subjected to a preparation procedure to allow for a morphological comparison among them. This preparation consisted of three basic steps. (1) 
In the first step, the Doppler shifts were removed by placing the line core of each intensity profile at the reference position $\Delta\lambda=0$ and the profiles were interpolated to a common wavelength grid. The line core position was determined  analytically as the minimum of a 
 Bezier fit \citep[e.g.,][]{bezier1968, farin2002}
around the minimum intensity point. 
All the polarization profiles were shifted and interpolated accordingly. (2) 
The second step involved signal normalization. All the intensity profiles were renormalized with respect to their continuum intensity $I_c$ after subtracting  their core intensity;
in contrast, all the Stokes $V$ profiles were renormalized with respect to their maximum unsigned value as $V(\lambda)=V(\lambda)/\text{max}(|V|)$. (3) Finally, we performed a polarity adjustment so that all the Stokes $V$ profiles display the same magnetic polarity (positive).

Also, while we considered all 14 observed wavelength points for the classification of the Stokes $I$ profiles, we consider only 13 wavelengths for the classification of the Stokes $V$ profiles, from -500 to 500 m{\AA}, that is, we neglect the continuum point at -700 m{\AA}, as its polarization signals are generally weak and of the order of the noise of the observations. Although taking this point into account  would not yield an important contribution to the morphology of Stokes $V$ profiles with a high signal-to-noise, it could have a more significant effect for the profiles with a lower signal-to-noise, therefore influencing their classification results.

\subsection{Selection of representative profiles}

We performed an exhaustive inspection of the Mg I b$_2$ data using the  CRisp SPectral EXplorer\footnote{CRISPEX is part of the IDL SolarSoft package, which is a set of integrated software libraries, databases, and system utilities that provide a common programming and data analysis environment for solar physics and can be accessed at \url{http://www.lmsal.com/solarsoft}.} 
 \citep[CRISPEX;][]{Vissers2012} tool to visualize and identify the different shapes of the Stokes $I$ and $V$ profiles present in our dataset.
We then attempted to identify the different classes of profiles in terms of their morphology by applying unsupervised clustering techniques, such as k-means \citep[e.g.,][]{MacQueen1967, Jain2010} and principal component analysis \citep[e.g.,][]{Wold1987, Jolliffe2016}.
However, we noticed that the unsupervised methods failed to identify all the different morphologies observed during the visual inspection conducted with CRISPEX, likely due to the fact that some of the differences between classes are rather subtle but also likely because the unsupervised methods were applied to a random selection of 10,000 profiles in order to speed up the computations.
Therefore, to avoid underestimating or overestimating the number of classes, we
 manually selected the representative profiles according to the CRISPEX visual inspection. 
In this process, we took extra care to avoid redundancy, and the representative profiles were chosen to capture the most distinctive signatures among the broad variety of shapes that were observed.  

For the classification of the Stokes $V$ profiles, we  considered only those pixels displaying signals above $5\sigma$, which correspond to $\sim 23\%$ of the total number of pixels in the FOV. Most of them  are associated with the magnetic regions. For the Stokes $I$ profiles, we employed all the pixels within the FOV.
Panels in Figure \ref{fig:4bef} show the pixels considered in each case on the continuum intensity maps normalized to the quiet Sun continuum intensity $I_c/I_{QS}$ and the location of the selected representative profiles, nine for the circular polarization, $V_R$, and 
16 for the intensity profiles, $I_R$.

Figure \ref{fig:4} shows the nine selected $V_R$ profiles before (top) and after (bottom) the preparation of the dataset as described in the previous subsection:
$V_R$ 0 is a normal profile displaying nearly anti-symmetric lobes; $V_R$ 1 is also a nearly anti-symmetric profile with wider lobes; $V_R$ 2 is a narrower profile with strong signals in the wings that increase in amplitude away from the line core; $V_R$ 3 is a narrow profile with strong signals in the wings displaying a convergent trend toward zero; $V_R$ 4 is also a narrow profile with 
non-zero wing signals but which remain rather constant as $|\Delta\lambda|$ increases; $V_R$ 5 is a narrow profile with negligible signals in the wings; $V_R$ 6 is a narrow profile with apparent extra-lobes given that the signal changes sign in the wings; $V_R$ 7 is a profile with a large positive amplitude asymmetry (i.e., the blue-lobe peak is stronger than the red-lobe peak); and $V_R$ 8 is a profile with a large negative amplitude asymmetry (i.e., the blue-lobe signal is weaker than the red-lobe signal).
Although we also found profiles that resemble single-lobed profiles (with either a blue or a red lobe only) as those commonly observed in typical photospheric lines, they are considered as extreme cases of classes 7 and 8 profiles, respectively.

Likewise, Figure \ref{fig:5} shows the 
16 selected $I_R$ profiles before and after the preparation of the dataset. 
While $I_R$ 0 closely resembles the average quiet Sun profile, $I_R$ 1 is slightly narrower, and $I_R$ 2 is slightly broader.
$I_R$ 3 and $I_R$ 4 both exhibit a normal line core but feature bright wings, likely due to emission, with the effect being more pronounced in the latter.
$I_R$ 5 displays a normal core but wider wings that might indicate some level of enhanced absorption. 
$I_R$ 6 is a narrow profile with bright wings, while $I_R$ 7 has normal wings but a narrower line core, likely due to emission in the inner wings. 
Likewise, $I_R$ 8 and $I_R$ 9 are very narrow profiles exhibiting emission across both wings, with a stronger effect in the latter where the emission surpasses the continuum intensity. 
Finally,  $I_R$ 10 to $I_R$ 15 display different types and degrees of wing asymmetries, which are somehow denoted by the line bisectors:  $I_R$ 10 shows emission in the inner blue-wing as well as in the outer red wing, $I_R$ 11 to $I_R$ 13 exhibit different degrees of blueward asymmetry, and $I_R$ 14 and $I_R$ 15 show varying levels of redward asymmetry.

In Sections \ref{sect:stokesV} and \ref{sect:stokesI}, the  Stokes $V$ and Stokes $I$ profile classes are further grouped into families based on shared morphological characteristics, physical properties, and spatial distribution.

\subsection{Classification of the profiles}

 \begin{figure*}[ht!]
   \centering
    
      \includegraphics[width=1\hsize]{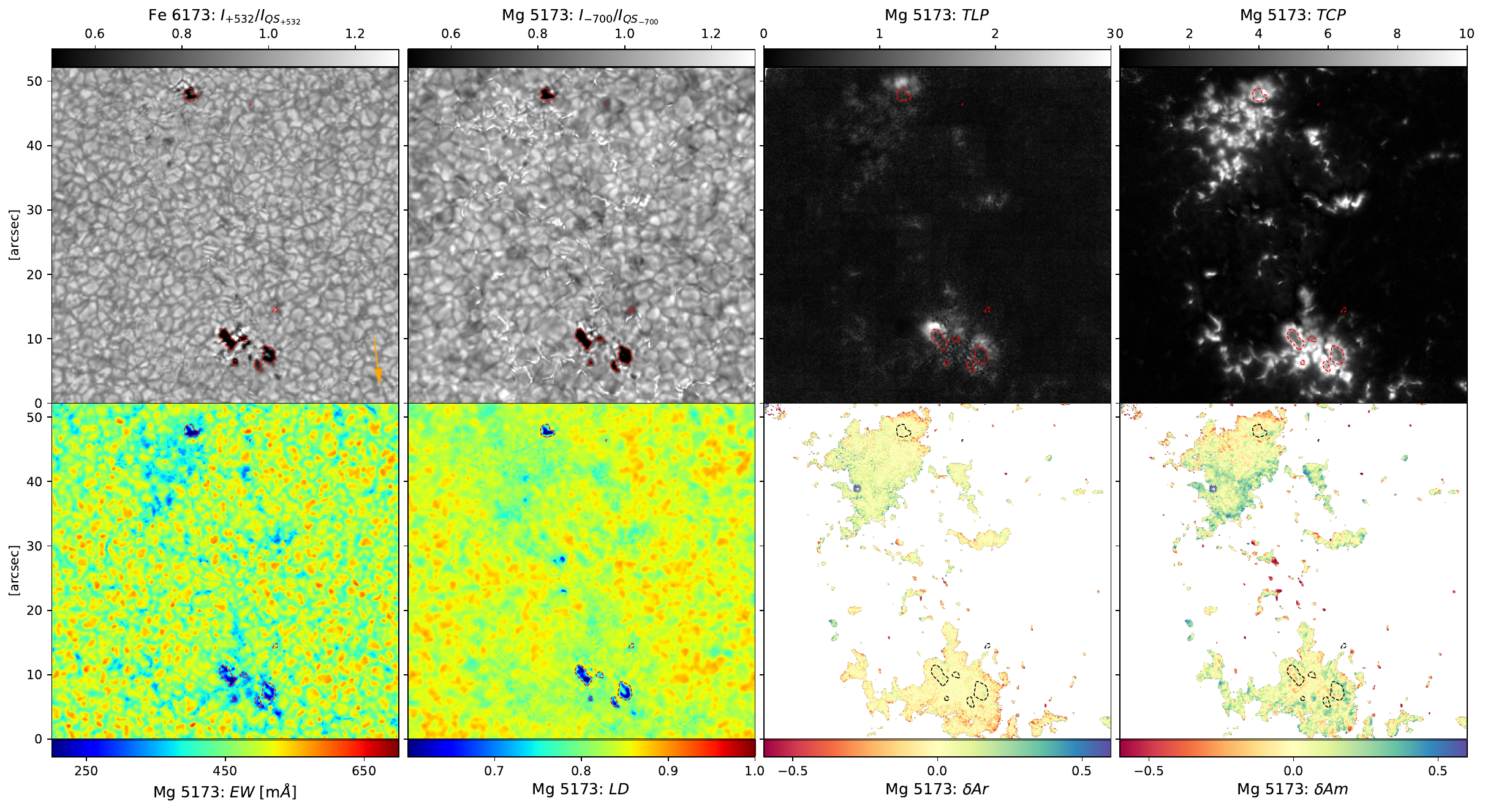} 
      \caption{Spectral and polarization properties of the Mg I b$_2$ line. Top: Reference continuum intensity in Fe  (at +532 m{\AA}) normalized to its mean QS value  $I_{+532}/I_{QS_{+532}}$,   continuum intensity in Mg (at -700 m{\AA} from the line core) normalized to its mean QS value  $I_{-700}/I_{QS_{-700}}$, and
      maps of 
      $TLP$ and 
      $TCP$. Bottom: 
      $EW$, 
      $LD$, 
      $\delta Ar$, and 
      $\delta Am$. The regions with Stokes $V$ signals smaller than $5\sigma$ have been masked in the maps of the Stokes $V$ asymmetries.
      Dashed contours are placed at $I_{-700}=0.5I_{QS_{-700}}$ of the Mg line to highlight the pore locations, and the orange arrow in the first panel points toward the disc center.
 }
         \label{fig:6mapsd}
   \end{figure*}

     \begin{figure*}[ht!]
   \centering
     
\includegraphics[width=1\hsize]{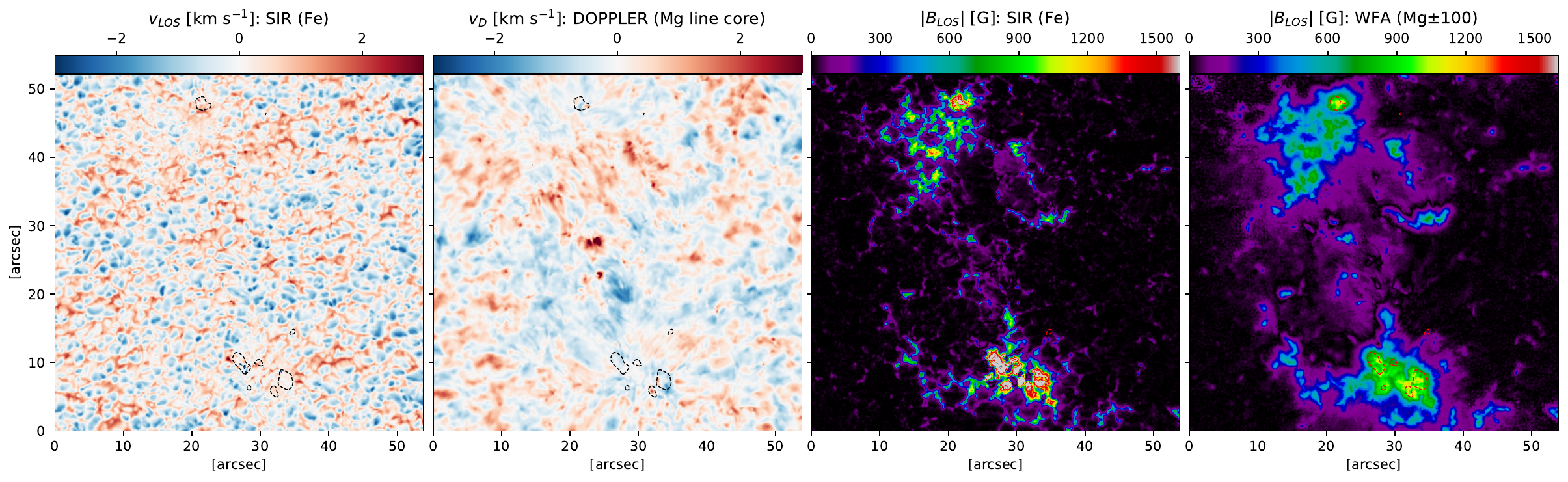} 
      \caption{Estimation of the LOS velocities and of the unsigned longitudinal magnetic field from classical methods. From left to right: $v_{LOS}$ from SIR inversions of the Fe line, $v_{D}$ from the Mg line core Doppler shifts, $|B_{LOS}|$ from SIR inversions of the Fe line, and  $|B_{LOS}|$ from the WFA applied to the Mg line core spectral region ($\pm100$ m{\AA}). The velocity maps were corrected from cavity errors.
 }
         \label{fig:6mapsd_VlosBlos}
   \end{figure*}

After the preparation of the profiles and the selection of the representatives, we computed the Euclidean distance, $eud$, of each profile in our dataset, $P_i$, with respect to the representative profile $P_{R}$ as

 \begin{equation}
   eud(P_i,P_R)=\sqrt{\sum_k(P_{i}(\lambda_k)-P_{R}(\lambda_k))^2},
   \end{equation} \label{eq:9}
   
 \noindent where $P$ is either Stokes $V$ or Stokes $I$, $k$ spans over $\Delta \lambda=\pm 500$ m\AA\ for the $V$ profiles and over the full wavelength range for the $I$ profiles; $i$ over the number of pixels considered in each case, and $R$ over the number of representatives.  Each profile $P_i$ was finally classified according to its closest representative, that is, with the $P_R$ that has the smallest $eud$.


\section{Analysis}

\subsection{Spectro-polarimetric proxies and classical inferences}

In order to analyze the properties of the different classes, we inspected some spectral and polarization quantities from the observed profiles, such as the total linear and circular polarization, $TLP$ and $TCP$ respectively, which were computed for a spectral window around the line core  (around $\pm100$ m{\AA}) as follows:

\begin{equation}
TLP=\int_{-0.1 \text{\AA}}^{+0.1 \text{\AA}}\sqrt{Q^2 (\lambda)+U^2 (\lambda)}d\lambda,
\label{eq:1}
\end{equation}

\begin{equation}
TCP=\int_{-0.1 \text{\AA}}^{+0.1 \text{\AA}}|V(\lambda)|d\lambda,
\label{eq:2}
\end{equation}

\noindent where  $Q$, $U$, and $V$ are normalized to the quiet Sun continuum intensity, $I_{QS}$; and $d\lambda= 50$ m\AA\ in the line core spectral window.

We also inspected the asymmetries of the Stokes $V$ profiles by computing their area and amplitude asymmetries, $\delta Ar$ and $\delta Am$ respectively, as follows:

\begin{equation}
\delta Ar=s\frac{\int_{-0.5\text{\AA}}^{+0.5 \text{\AA}} V(\lambda)d\lambda}{\int_{-0.5 \text{\AA}}^{+0.5 \text{\AA}}|V(\lambda)|d\lambda},
\label{eq:3}
\end{equation}

\begin{equation}
\delta Am=V_b-Vr.
\label{eq:4}
\end{equation}

\noindent In Equation \ref{eq:3}, the symbol $s$ refers to the sign of the Stokes $V$ blue lobe. In Equation \ref{eq:4}, $V_b$ and $V_r$ are the unsigned amplitudes of the blue and red lobe, respectively, normalized to the maximum signal of Stokes $V$. 
In this way, positive values of $\delta Ar$ ($\delta Am$) indicate a larger area (amplitude) in the blue lobe than in the red lobe and vice versa.

From the observed intensity profiles, we computed LOS velocities  by considering the Doppler shift of the line core, $\Delta \lambda_{lc}$:

\begin{equation}
v_{D}=\frac{c \Delta \lambda_{lc}}{\lambda_0},
\label{eq:5}
\end{equation}

\noindent where $c$ is the speed of light, $\lambda_{lc}$ is the line core wavelength position determined by the analytic minimum of a Bezier fit. 
The reference wavelength for zero velocity is taken from the average intensity profile in the dataset (see Paper I), and $\lambda_0=5172.68$ \AA\ for the Mg line.

The line bisectors $I_{bis}$ were calculated at different intensity levels using linear interpolation of the observed profiles.
In addition, we obtained the equivalent width $EW$ and the line depth $LD$ of the intensity profiles as follows:

\begin{equation}
EW=\int_{-0.5\text{\AA}}^{+0.5 \text{\AA}} (1-I(\lambda)/I_c)d\lambda,
\label{eq:6}
\end{equation}

\begin{equation}
LD=1-(I_{min}/I_c),
\label{eq:7}
\end{equation}

\noindent where $I_c$ refers to the continuum intensity, and $I_{min}$ to the minimum intensity value in the profile.

We computed the longitudinal component of the magnetic field by applying the weak-field approximation \citep[WFA;][]{Deglinnocenti1973} in three different spectral windows within the Mg line: $\pm500$ m{\AA}, $\pm300$ m{\AA}, and $\pm100$ m{\AA}, which progressively sense higher layers in the solar atmosphere (see Paper I for a more detailed discussion). Thus, this method provides information about the gradients with height of the magnetic field. 
Finally, we performed a one-component LTE inversion of the Fe I 6173 \AA\ line in the full dataset using the Stokes Inversion based on Response functions \citep[SIR;][]{Ruiz1992} code in order to get information about the physical parameters in the photosphere for the observed FOV. The detailed configuration of these inversion are described in Paper I. 

\subsection{NLTE inversions of representative profiles}

We used the Departure coefficient aided Stokes Inversion based on Response functions \citep[DeSIRe;][]{Ruiz2022} code to perform  inversions of the four Stokes profiles simultaneously in the Fe I 6173 \AA\ line (which is treated in LTE) and the Mg I 5173 \AA\ line (treated fully in NLTE) for some representatives of the Stokes $I$  and of the Stokes $V$ classes  of Mg in order to compare the resulting stratifications with those inferred with the classical methods, as well as to get information on the temperature in the low chromosphere. 
 For simplicity, we refer to these two-line inversions as NLTE inversions. However, we emphasize that only the Mg line is treated in NLTE, while the Fe line is inverted under LTE conditions.   
 
Since the Mg pseudo-continuum wavelength at $\Delta\lambda=-700$ m\AA\ is not an actual continuum wavelength, the inverted representative profiles were previously normalized to a continuum intensity value estimated from the Fourier Transform Spectrometer (FTS) atlas \citep{Neckel1984}. 
The inversions used the 13-level atomic model of Mg \citep{Quintero2018} and
applied opacity fudge \citep{Bruls1992} while inverting all cycles under a full NLTE treatment (approach that proved to perform better for the Mg I b$_2$ line in the tests presented in Paper I despite being computationally slower).

We are interested in understanding the physics producing the different shapes of the representative profiles. 
However, we could not find a general initial inversion setup that would properly reproduce the distinctive features of all the representative profiles.
Thus, the inversions were made separately according to the individual needs. 
Through trial and error we found the number of cycles, weight distribution for the Stokes parameters, and the number of nodes in the atmospheric parameters that provide reasonably good fits to 
the selected representatives.
 Such configurations are indicated in Tables \ref{tab1} and \ref{tab2}  for the $V_R$'s and $I_R$'s, respectively. 
 While we inverted all nine sets of Stokes profiles for the $V_R$'s, we only inverted 13 out of 16 sets for the $I_R$'s, although ensuring that all Stokes $I$ families are represented.

In most of the cases, the inversions were performed by assigning larger weights to Stokes $I$ and $V$ than to Stokes $Q$ and $U$, as the latter are generally near the noise level, particularly for the Mg line. Likewise, in most of the cases the inversion initial setup considers a simple one-component model atmosphere (i.e., $\mathit{ff}=1$ and the stray light is assumed to be zero) whose temperature stratification between photosphere and chromosphere is based on the FALC model \citep{Fontela1993}. The exceptions are the set of profiles corresponding to 
$I_R$ 10 to $I_R$ 15,
 and those for $V_R$ 7 and $V_R$ 8, whose asymmetries could only be fitted using a two-components atmosphere with different initial magnetic field and velocity values. 
The spectral point-spread-function (PSF) of the CRISP prefilters are taken into account during the inversions by using Gaussians with a full width at half maximum of 25 and 50 m\AA\ for the Mg and Fe lines, respectively.

  Although the core of the Mg I b$_2$ line is most sensitive to the low chromosphere, our input model extends beyond that, up to $\log(\tau)=-6.2$.  
Nonetheless, the resulting stratifications can only be trusted within the line sensitivity height range --- see response functions (RFs) in Figure \ref{fig:15rfs} which peak near $\log(\tau)=-4.3$ in the core of the Mg 5173 line and do not extend beyond $\log(\tau)=-5$.
Finally, error bars are not provided by DeSIRe. Thus, we calculated the uncertainties by iteratively adding random noise to the observed profiles (of the same order as the estimated noise in the observations) and performing 100 inversions of these profiles with the exact same input model. The standard deviation given by the 100 retrieved atmospheres are taken as an estimation of the uncertainties of the parameters at each optical depth.


\section{Results}

Figure  \ref{fig:6mapsd} shows normalized continuum intensity maps for the Fe line (at +532 m{\AA} from its core) and for the Mg line (at -700 m{\AA} from its core). The maps display noticeable differences, such as the size of the granules being smaller in the Fe map and the presence of bright structures or plage in the Mg map that appear with smaller sizes or are completely absent in the Fe map. These differences are due to the fact the wavelength point at -700 m{\AA} from the Mg line core still lies within the far blue-wing of the spectral line and is not an actual continuum value, although we refer to it as such for simplicity.
Therefore, the Mg continuum map actually portrays a higher layer of the solar atmosphere, while the Fe map shows the low photosphere.

The spectro-polarimetric proxies computed for the Mg line are displayed in Fig.\ \ref{fig:6mapsd}. They show that the pores and the bright magnetic structures observed in the FOV exhibit considerably larger $TCP$ values than  the quiet Sun granulation. In contrast, the $TLP$ does not stand well above the noise over most of the FOV. However, there is still some  linear polarization signal near the pores, 
observed over an asymmetrical canopy-like region around the pores, which is stronger and more extended toward the limb side. This result is understood to be due to a projection effect, given that the observations were made at a moderate heliocentric angle.
  The surrounding plage regions also contain strong signals of circular polarization, while faint traces of $TLP$ are detected in these sites.

The maps of $\delta Ar$ and $\delta Am$ display generally lower asymmetries in the pores and the centers of the bright structures, but they show an overall increase toward the edges of those magnetic regions, probably reflecting the presence of canopy-like structures around them \citep{Martinez2012b}. 
Fig.\ \ref{fig:6mapsd} also reveals that the Mg I b$_2$ line exhibits substantially narrower and shallower intensity profiles within the magnetic regions (with $EW < 400$ m\AA\ and $LD < 0.8$) compared to the quiet Sun. The granules show wide and deep intensity profiles (with $EW > 500$ m\AA\ and $LD > 0.85$), while the profiles are slightly narrower in the inter-granular regions ($EW \sim 450$ m\AA\ and $LD \sim0.83$). 

 \begin{figure}[h!]
   \centering        
 \includegraphics[width=0.7\hsize]{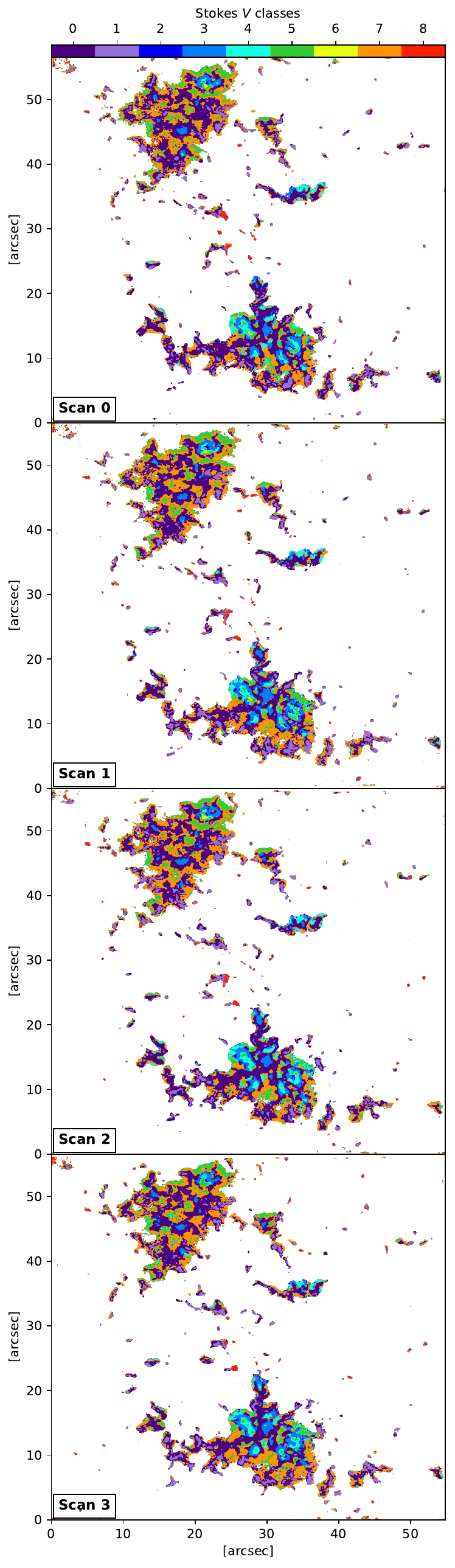} 
      \caption{Spatial distribution of the nine classes of Stokes $V$ profiles of the Mg I b$_2$ line found in the four scans of our dataset according to the $eud$ classification. 
      The images show the full FOV.
              }
         \label{fig:6}
   \end{figure}

The Doppler velocity map of Mg, displayed in Figure \ref{fig:6mapsd_VlosBlos},  showcases a flow structure typical of the high-photosphere/low-chromosphere, featuring strong flows directed toward the pores along canopy-like fibrils. Additionally, the figure presents the LOS velocities resulting from the SIR inversions of the Fe 6173 line. These results depict the convective granulation pattern characteristic of the lower photosphere, with upflows in the centers of the granules and downflows
in the intergranular lanes.

The unsigned photospheric and chromospheric longitudinal magnetic field maps from SIR and the WFA, respectively, are also displayed in Fig.\ \ref{fig:6mapsd_VlosBlos}. These maps reveal that the strongest magnetic fields in the FOV are concentrated inside the pores, 
although the bright structures in the plage region are also moderately magnetized. Notably, these magnetic regions all appear more compact in the photosphere and exhibit an enlarged area in the high-photosphere/low-chromosphere, indicating a rapid opening of the magnetic field with height \citep[e.g.,][]{Buehler2015}.

\section{Stokes $V$ classification} \label{sect:stokesV}

Figure \ref{fig:6} presents the spatial distribution of the nine classes of Stokes $V$ profiles identified in our Mg data, based on the $eud$ classification. The distribution of these classes remains remarkably consistent across the four temporal scans, demonstrating the robustness of the Euclidean distance classification method. 
This consistency is expected, as the main magnetic structures (such as pores and plage) are unlikely to undergo significant changes within the short temporal scales of our observations (approximately 2 minutes between scans). Therefore, the shape of the Stokes $V$ profiles and the class assignments should not vary substantially between scans, unless influenced by very fast-evolving processes.

Furthermore, as shown in the first column of Figure \ref{fig:7}, the mean profiles of the classes (red lines) closely resemble the corresponding representatives (blue lines) and preserve the main spectral features. This further validates the robustness of the $eud$ classification.
This consistency holds even in cases where there is a significant variation among the individual profiles within the same class, as indicated by the shaded areas for classes 2, 6, and 8. These classes are also the smallest groups, respectively representing $0.40\%$, $0.20\%$, and $1.86\%$ of the total pixel sample (see Table \ref{tab_res1}).
However, the method 
does not perfectly capture the spectral signatures of some profiles. For instance, some of the farthest profiles in class 8 show a positive $\delta Am$, unlike the representative, which is characterized by a negative $\delta Am$. Nonetheless, these profiles represent only a small fraction of the class, as depicted in the distribution of amplitude asymmetry in the last column of Fig.\ \ref{fig:7}.

   \subsection{Family A}
\begin{figure*}[!htbp]
   \centering
          
        \includegraphics[width=0.9\hsize]{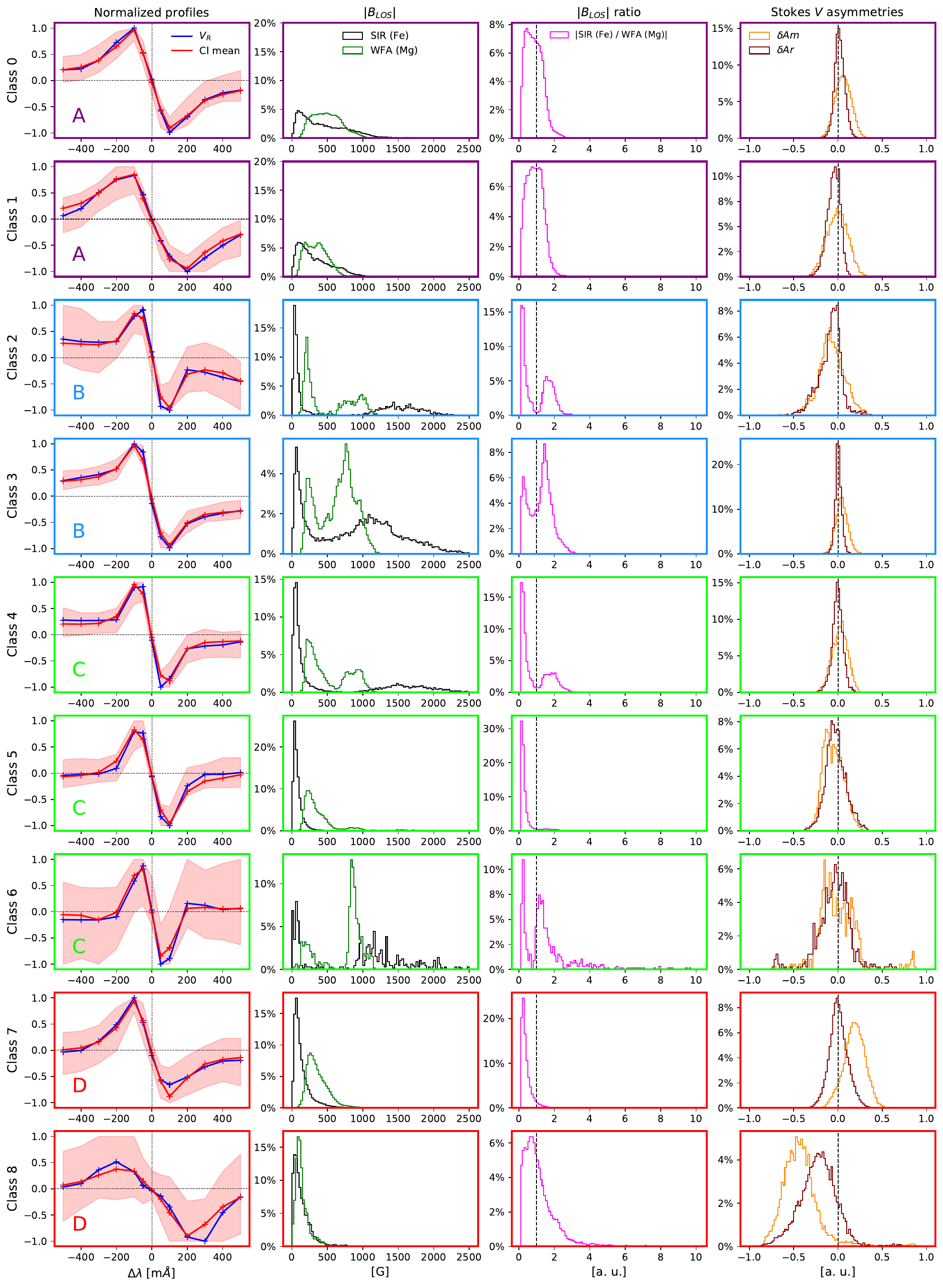}
      \caption{Polarization and magnetic properties of the Mg I b$_2$ Stokes $V$ classes.  
      From left to right: 
      Representative profiles $V_R$ (blue) and  mean class profiles 
      (red), with shaded areas indicating the range of variation of the profiles within the classes; histograms of the 
      $|B_{LOS}|$ distribution from the SIR inversions (black) and from the WFA applied to the core of the Mg line (green); 
       the photospheric to chromospheric $|B_{LOS}|$ ratio; 
and the amplitude (orange) and area (maroon) asymmetries. 
    Letters in column 1 and frame colors highlight different families. 
              }
         \label{fig:7}
   \end{figure*}

\begin{figure*}[ht!]
   \centering

        \includegraphics[width=\hsize]{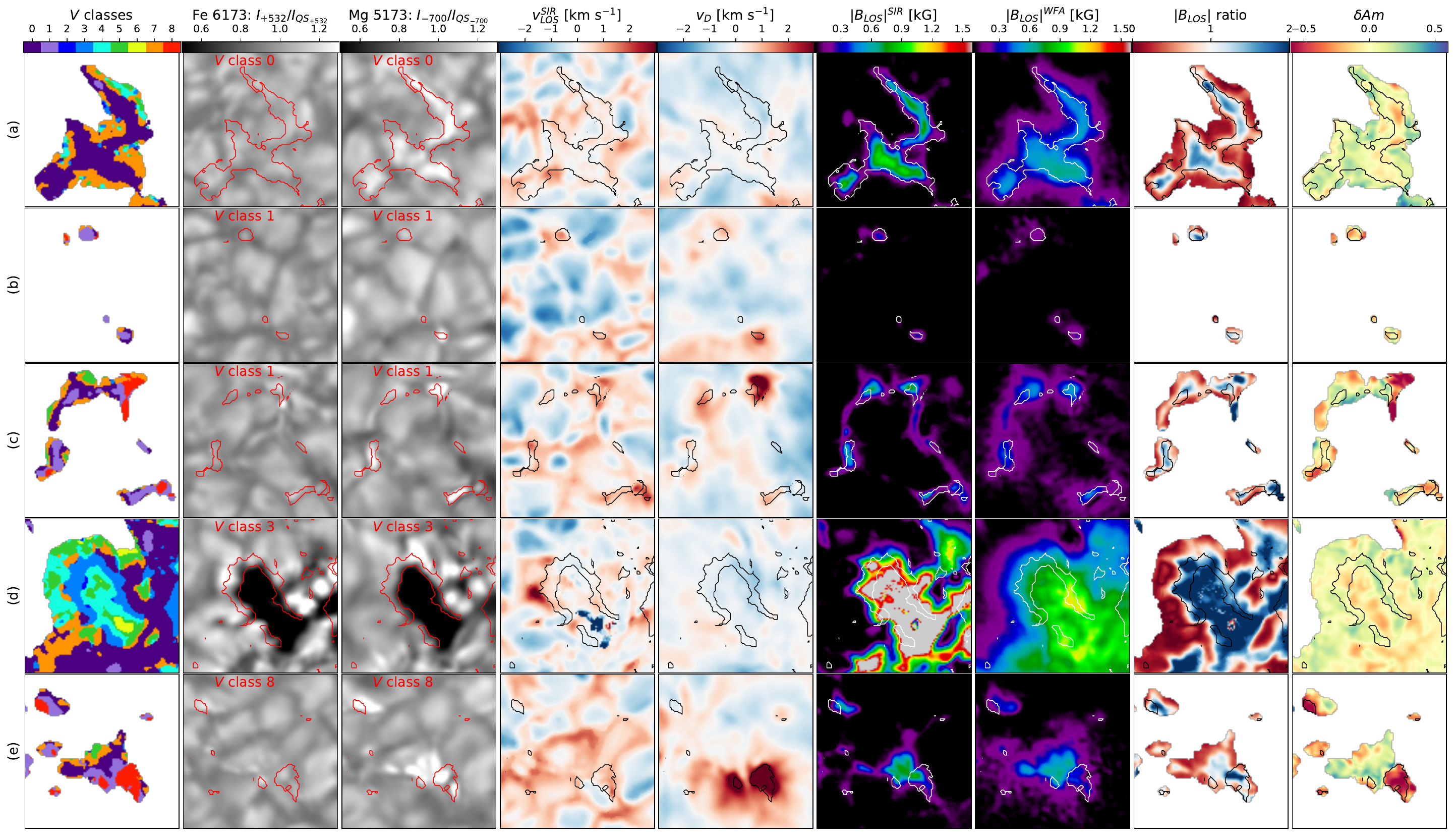} 
       \caption{Examples of different structures and their associated Stokes $V$ classes and physical properties.  From left to right:  $eud$ classification of Mg Stokes $V$,  Fe continuum intensity (at +532 m{\AA}) normalized to its mean QS value, Mg continuum intensity (at -700 m{\AA}) normalized to its mean QS value, photospheric LOS velocity from the SIR inversions of the Fe line, Doppler velocity of the Mg line core, unsigned LOS magnetic field from the SIR inversions and from the WFA in the Mg $\pm100$ m\AA\ spectral window, the  photospheric to chromospheric $|B_{LOS}|$ ratio, and the amplitude asymmetries of the Mg Stokes $V$ profiles. All the examples show a $\sim 6\arcsec \times 6\arcsec$ sub-field. For each row, the contours enclose the $V$ class specified in the intensity maps.
              }
              
         \label{fig:8}
   \end{figure*}

   As indicated in Table \ref{tab_res1}, the largest group corresponds to class 0 (dark purple in Figure \ref{fig:6}), with approximately $40\%$ of the selected profiles falling into this category. 
   Profiles in class 0 span over a broad range of  magnetic field strengths, as shown by the  
   histograms in the second column of Fig.\ \ref{fig:7}, each portraying a different height in the atmosphere (Paper I). 
  The photospheric field (black distribution) covers a  range of values larger than the chromospheric magnetic field (green) and seems to display a bimodal distribution, with a population of weak magnetic fields peaking at around $100$ G, and a second population at moderate field strengths (from approximately $400$ G, where the distribution presents a rebound) displaying a tail that extends beyond kiloGauss values. Such bi-modal behavior is observed for most of the classes in both the photospheric and chromospheric magnetic field distributions, except in classes 7 and 8 where only one peak is distinguishable. The two magnetic field populations are largely separated in the classes from 2 to 6, by more than 1000 G in the photosphere, while for the chromospheric field this separation is smaller (see also Table \ref{tab_res1}). In classes 0 and 1, the separation between the two populations of the chromospheric field is presumable very small, making them barely distinguishable.
These results indicate that, regardless of the profiles' amplitude scale,  the morphology of Stokes $V$ can be similar across different magnetic field regimes, allowing for a given Stokes $V$ class to be observed in various solar situations. 
This is shown in the examples of Figure \ref{fig:8}. 
Class 0 profiles in example (a) are located within  the core of a bright magnetic patch, they surround class 1 profiles in a pair of magnetic points in (b), they  spread over the plage near a pore in (d), and around class 1 profiles in  bright magnetic structures of irregular shapes, such as in examples (c) and (e). These examples display different magnetic field values at the location of class 0 profiles. However, they all share common characteristics,  including moderate-to-low velocity and magnetic field gradients between the photosphere and chromosphere (with a $|B_{LOS}|$ ratio around 1), mild-to-zero Stokes $V$ asymmetries, and the presence of bright magnetic structures in the Mg intensity maps. This excess brightness is often not observed in the Fe intensity maps or appears with lower contrast.

   Class 0 profiles 
    frequently occur alongside and around the less abundant class 1 profiles (approximately $10\%$ of the sample, as shown in Table \ref{tab_res1}). Both classes, 0 and 1,  are composed mainly of anti-symmetric profiles with low $\delta Am$ and $\delta Ar$ values, 
   and display strikingly similar distributions in Fig.\ \ref{fig:7}, a characteristic that allows us to categorize these two classes into a single family for further simplification (Family A). There are subtle differences though, such as the width of the lobes being larger for class 1 profiles. Additionally, although the peak of the photospheric to chromospheric $|B_{LOS}|$ ratio distribution is located near 1 in both classes (indicating nearly constant fields), it is slightly shifted toward values greater than $1$ in class 1. This shift indicates a dominance of magnetic fields that decrease with height, as expected to occur in the central core of the magnetic structures where class 1 profiles are largely observed.

Figure \ref{fig:15} presents the results of the NLTE inversions of the Stokes $V$ representatives, along with the estimated uncertainties.
Generally, the uncertainties in temperature, velocity, and magnetic field increase with height, as expected, due to the lack of sensitivity above $\log(\tau)=-5$, as indicated by the response functions in Figure \ref{fig:15rfs}. 
However, in cases requiring a larger number of free parameters (see Table \ref{tab1}), the uncertainties can be significant at all optical depths.

\begin{table*}[ht!]
\centering
\caption{Summary Table of Fig.\ \ref{fig:7}. 
}
\begin{tabular}{ c c c c c c c c c c    } 
\hline
$F$ & Class & $\%$ pix & mean  &  mean  & mean  & mean & mean  & mean  &  mean   \\
&  &   & ($|B_{LOS}^{SIR}|$) & ($|B_{LOS}^{WFA}|$) &($|B_{LOS}|$ ratio) &($v_{LOS}^{SIR}$) & ($v_{LOS}^{DOP}$) & ($\delta Am$) & ($\delta Ar$) \\
 &   &  & [G] & [G] & & [km s$^{-1}$] & [km s$^{-1}$] &  &  \\
\hline
\multirow{2}{*}{A} & 0 & 40.27 & 475
& 495
& 0.97
  & 0.48 & 0.14 & 0.06 & 0.02  \\ 
 & 1 &  9.57 & 333
 & 350
  & 1.01
   & 0.51 & 0.53 & -0.01 & -0.05  \\ 
\hline
\multirow{2}{*}{B} & 2 & 0.40 & 714/72/1540 & 522/228/881 & 0.94/0.28/1.75   & 0.03 & 0.13  &  -0.07 & -0.09  \\ 
 & 3 &  6.41 & 876/163/1242 & 636/301/783 & 1.20/0.48/1.63   & 0.62 & 0.04  & 0.04 & 0.00  \\ 

\hline
\multirow{3}{*}{C} & 4 &  4.11 & 631/97/1687 & 530/298/845 & 0.83/0.28/1.91   & 0.26 & -0.05  & 0.02 & 0.00  \\ 
  & 5 &  8.32 & 197/79/1794 & 354/289/725 & 0.37/0.24/2.04 & -0.03 &-0.03 &  -0.05 & -0.02  \\ 
 & 6 &  0.20 & 1310/82/1848 & 725/239/896 & 1.53/0.35/2.18  & 0.29  & 0.03 &  -0.01 & -0.04  \\ 
\hline
\multirow{2}{*}{D} & 7 &  28.85 & 151 & 359 & 0.39  & 0.13 & 0.09 &  0.19 & 0.01  \\ 
 & 8 &  1.86 & 137 & 139 & 1.49    & 0.52 & 1.73 &  -0.42 & -0.22  \\ 
\hline
\end{tabular}
\label{tab_res1}
\tablefoot{From left to right, columns indicate the family $F$, the class,  the percentage of pixels in each class relative to the total pixel sample (recalling that only  Stokes $V$ profiles with signals above $5\sigma$ were classified, i.e, $\sim23\%$ of the dataset), and the mean values in the histograms of Fig.\ \ref{fig:7} for each class. For the classes clearly exhibiting two distinct populations in the histograms, three mean values are provided: one for the entire distribution and one for each population.}
\end{table*}

The NLTE inversions for the representatives of Family A, $V_R$ 0 and $V_R$ 1 (whose locations in the FOV are specified in Fig.\ \ref{fig:4bef}), result in temperature profiles that are substantially hotter than the FALC model above $\log(\tau)=-0.7$, the height corresponding to the maximum response of Stokes $I$ to changes in temperature within the core of the Fe line. Below this height, the temperatures are close to the FALC profile, since the Fe continuum intensities take values around $I_{QS}$ in these profiles, as in most bright structures seen in Mg as mentioned above. 
The height stratification of $v_{LOS}$, $B$, and $\gamma$ remains fairly constant for $V_R$ 0, while $V_R$ 1 shows
 a small gradient in the magnetic field. However, the magnetic field values for $V_R$ 1 are similar at the heights of maximum response of the cores of the Mg and Fe lines to changes in the magnetic field ($\log(\tau)=-4.3$ and $\log(\tau)=-0.8$, respectively), with values near 500 G. 
As indicated in the corresponding panels of the figure, the estimations made with the classical methods for $V_R$ 0 and $V_R$ 1 
 are in agreement, within the errors, with the NLTE inversion results for the velocity and the magnetic field configuration, except for the SIR inversions inferring magnetic field inclinations that are about 20 degrees more vertical than those inferred by DeSIRe in the low photosphere.

   \subsection{Family B}
  Another family is formed by classes 
 2 and 3 (Family B). 
  Both classes consist of narrow profiles with strong signals in the wings of Stokes $V$ whose shapes are 
  either diverging or converging.
   They are clearly associated with two very different magnetic regimes, as indicated by their $|B_{LOS}|$ histograms in Fig.\ \ref{fig:7}. Although they mostly appear agglomerated 
   inside and around the periphery of the pores, they also occur  
   further out in regions with weaker magnetic fields (see different examples in Fig.\ \ref{fig:8}). This dual occurrence also explains the double-peaked distributions in the photospheric to chromospheric $|B_{LOS}|$ ratio histograms. The peak at values greater than $1$ corresponds to the pores and their periphery, where the magnetic field decreases with height (see example (d) in Fig.\ \ref{fig:8}), while the peak at values less than $1$ corresponds to regions outside the pores, where the magnetic field increases with height due to the opening of the magnetic field around the pores.
   The asymmetries of the profiles  and the inferred velocity gradients are also small for this family. However, the strong wing signals in class 2 profiles produce a moderate negative tail in the $\delta Am$ and $\delta Ar$ distributions.

 \begin{figure*}[!htbp]
   \centering

\includegraphics[width=0.98\hsize]{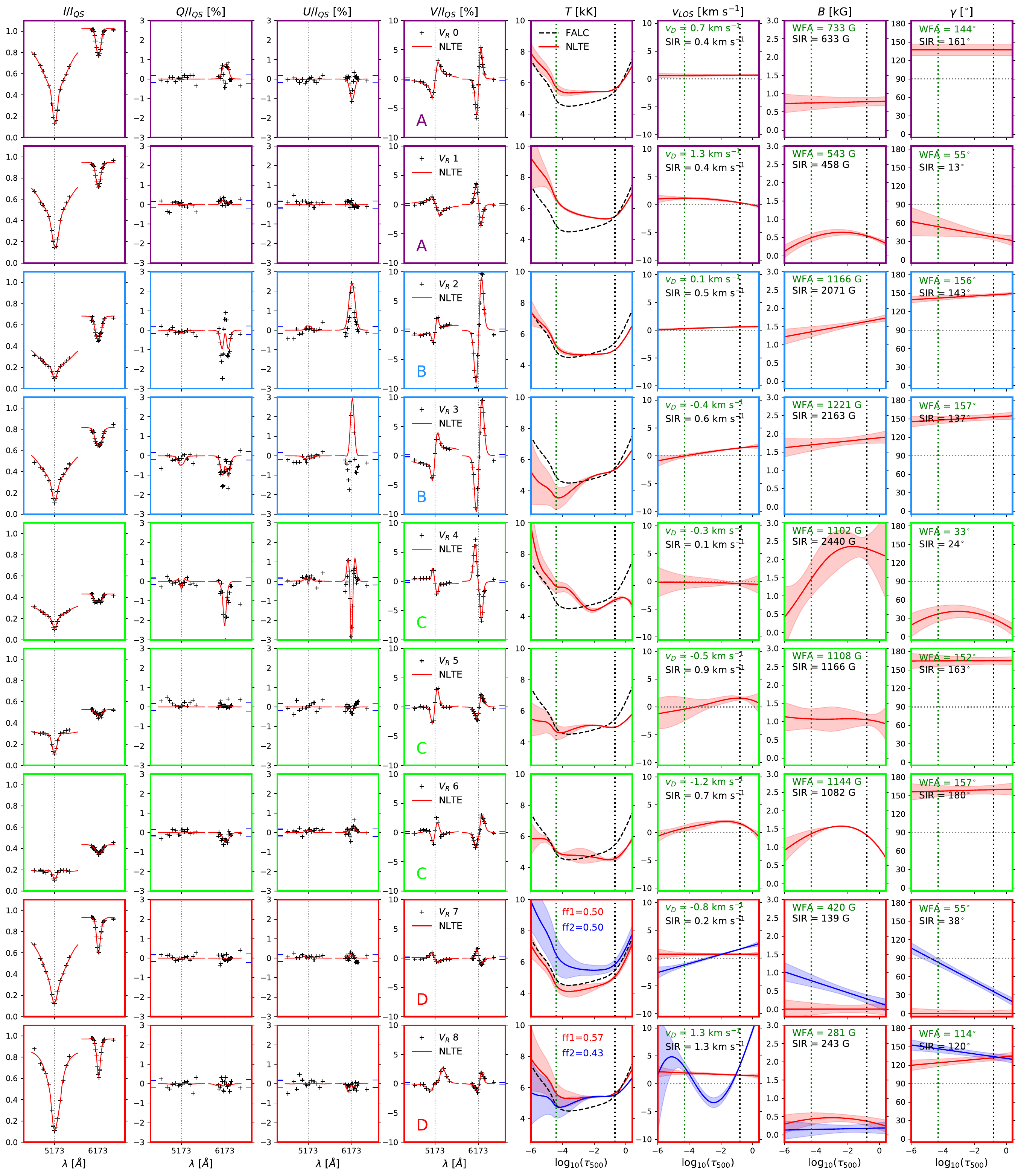} 
      \caption{NLTE inversions of Stokes $V$ representatives. 
      From left to right: Observed Stokes profiles (black markers) in the pair of Mg I 5163 and Fe I 6173 lines normalized to their mean QS continuum intensity (the Mg continuum intensity was estimated from the FTS atlas profile), 
       and their best fits (red lines). The blue tick-marks in the $Q/I_{QS}$, $U/I_{QS}$, and $V/I_{QS}$ panels indicate the $\sigma$ levels for each spectral line.  
       The resulting temperature stratifications are shown in the fifth column (red lines), overlaid on the initial FALC model (black dashed line), followed by the retrieved  LOS velocity, magnetic field intensity, and magnetic field inclination. 
       The blue lines in the last two rows show the stratifications for a second atmospheric component, with filling factors $\mathit{ff}$ indicated in the temperature panel. The uncertainties are indicated by the shaded areas. For comparison, labels in the last three columns indicate the results obtained with the classical methods. 
      Vertical green and black dotted lines in the panels of $T$, $v_{LOS}$, $B$, and $\gamma$ indicate the height of maximum response to changes in the given parameter within the Mg and Fe lines core, respectively, according to the response functions displayed in Figure \ref{fig:15rfs} (these heights are only valid for $V_R$ 0, but are shown as a reference for all representative profiles). Letters in column 4 and frame colors highlight the different families.
              }
         \label{fig:15}
   \end{figure*}

%
%

As shown in Fig.\ \ref{fig:4bef}, the representative profiles for  classes 
2 and 3
 stem from the periphery and the interior of pores. 
These regions exhibit the strongest $TLP$ of the Mg line in the observed FOV (see Fig.\ \ref{fig:6mapsd}). However, the Stokes $Q$ and $U$ signals are not well above the noise level even in these regions.  This is shown for the representative profiles 
$V_R$ 2 and $V_R$ 3
in Fig.\ \ref{fig:15}, which display linear polarization signals in Mg near the noise level, although they do show strong linear polarization signals in the Fe line. 
The NLTE inversions  of these profiles yield temperature stratifications that are substantially cooler than the FALC model, particularly in the lower layers of the atmosphere (see Fig.\ \ref{fig:15}). This aligns with their low intensities and shallow Stokes $I$ profiles. The results also indicate that the temperatures exceed the FALC model higher up for $V_R$ 2, above $\log(\tau)\sim -2.5$, 
 unlike $V_R$ 3 which displays very low temperatures above this height, where the estimated uncertainties become very large. 
The results for the magnetic field strength differ from those of the WFA and the SIR inversions by a few hundred Gauss. 
Nonetheless, both classical and NLTE results are qualitatively in agreement,  as they both estimate rather oblique kiloGauss magnetic fields in the photosphere and in the chromosphere, with the field strength decreasing with height while displaying fairly constant field inclinations and mild LOS velocities.

\subsection{Family C}
Family C consists of classes
4, 5, and 6. 
These classes all exhibit narrow profiles with generally weak signals in their wings, with class 4 exhibiting nearly constant although non-zero wing signals, class 5 showing nearly zero wing signals, and with class 6 profiles displaying small extra lobes of opposite polarity. 
 These classes are found 
  inside and outside the pores, as illustrated in example (d) of Fig.\ \ref{fig:8}. 
  Therefore, both classes display double-peaked distributions for $|B_{LOS}|$ and the $|B_{LOS}|$ ratio, as shown in Fig.\ \ref{fig:7}. However, the second population is very small for class 5, indicating that this class is more common outside the pores. In contrast,  class 6 profiles are mostly found inside pores and, as indicated in Table \ref{tab_res1}, they exhibit the strongest decrease in the magnetic field strength with height  among all the classes (average $|B_{LOS}|$ ratio $=2.18$, but reaching values as high as $27$ in some regions within the pores' center).
  The presence of strong extra lobes in some class 6 profiles produces large asymmetries, as shown by the $\delta Am$ and $\delta Ar$ histograms. Additionally, class 6 profiles are typically associated with very shallow intensity profiles that exhibit some emission in the wings. These characteristics might affect the estimation of the LOS velocities with the employed methods and could be  responsible for the large velocity gradients observed in example (d) of Fig.\ \ref{fig:8} for class 
  6 inside the pore.

The representative profiles  
$V_R$ 4, $V_R$ 5, and $V_R$ 6
come from the interior of a pore, as indicated in Fig.\ \ref{fig:4bef}, so that they 
all display low continuum intensity and very shallow Stokes $I$ profiles. 
The NLTE inversions shown  in Fig.\ \ref{fig:15}  retrieve low temperatures in the photosphere for all three representatives, reaching values down to 5200 K, 4900 K and 4500 K, respectively. However, while the temperature profile of $V_R$ 4 exceeds the FALC model  above $\log(\tau)\sim -2.5$, the temperature profiles of $V_R$ 5 and $V_R$ 6 remain rather flat,  although they show a small temperature increase near $\log(\tau)=-2$ that overcomes the FALC model up to $\log(\tau)\sim -4$ to reproduce  the apparent emission observed in the inner wings of their Mg intensity profiles.
The NLTE inversions infer largely oblique and vertical magnetic fields, respectively, with intensities above 1 kG that either decrease with height or
remain nearly constant 
between the heights of maximum response of the core of the spectral lines, and low-to-moderate LOS velocities. 
Such velocities and magnetic field configuration are compatible with those inferred by the classical methods for $V_R$ 4 and $V_R$ 5, whose results fall within the estimated uncertainties of the NLTE inversions. However, the methods differ by a few hundred Gauss in the case of $V_R$ 6 and show a discrepancy in the estimated $v_{LOS}$ of about 1 km s$^{-1}$.

\subsection{Family D}
Family D is composed of classes 7 and 8, both containing asymmetric Stokes $V$ profiles. 
Class 7 profiles, which exhibit positive amplitude asymmetry, are predominantly found at the edges of bright magnetic structures. This is illustrated in examples (a), (b), (c), and (e) of Fig.\ \ref{fig:8}, where they appear surrounding classes 0 and 1. Additionally, class 7 profiles are observed in some areas outside the pores, as shown in example (d).  
  As indicated in Table \ref{tab_res1}, class 7 is the second-largest group, accounting nearly $29\%$ of the profiles in the sample. The distribution of  amplitude asymmetry for class 7 in Fig.\ \ref{fig:7}  shows a clear tendency toward large positive values, as expected. 
  In contrast, the area asymmetry distribution is nearly zero-centered because these profiles typically display a blue lobe with larger amplitude than the red lobe, while the red lobe tends to have a larger width, reducing the area asymmetries.
  The magnetic field distributions for class 7 show a single peak and cover larger field strengths in the chromosphere than in the photosphere,  with tails that do not extend beyond 1 kG. 
The $|B_{LOS}|$ ratio is less than $1$ for the majority of the profiles in this class, confirming that the longitudinal magnetic field increases with height around the magnetic structures. 
These results are consistent with a low-lying magnetic canopy that forms in the high-photosphere/low-chromosphere, indicating a rapid opening of the magnetic field with height at the edges of bright magnetic structures.  This is not surprising since such a magnetic configuration gives rise to strong magnetic field gradients along the LOS, which has been proven to favor the formation of amplitude asymmetries in the circular polarization of some photospheric spectral lines \citep{Martinez2012b}, similar to
 the asymmetric Stokes $V$ profiles of the Mg line in class 7. 
 In the vicinity of pores, class 7 profiles appear 
  further out from the magnetic core compared to the bright structures. This might indicate a difference in the height at which the magnetic canopy is formed, suggesting that it is placed at higher layers in the case of the pores due to the stronger magnetic fields they host.

Class 8 profiles, which exhibit negative amplitude asymmetry, also display a distinct distribution of $\delta Am$ and $\delta Ar$ in Fig.\ \ref{fig:7}, with a clear tendency toward large negative values, as expected. 
These profiles are much less abundant than those in class 7, as indicated in Table \ref{tab_res1}, and are typically found in the core or at the edges of  bright magnetic structures that display strong downflows in the chromosphere, as illustrated in examples (c) and (e) of Fig.\ \ref{fig:8}. They show a single peak in the  $|B_{LOS}|$ distributions, with tails that do not extend beyond 1 kG. The $|B_{LOS}|$ ratio distribution peaks near 1 and exhibits an extended tail toward larger values, indicating  the presence of strong gradients where the magnetic field decreases with height, as in the core of magnetic structures.
  Although not depicted in Fig.\ \ref{fig:7}, the $v_{LOS}$ and $\Delta v_{LOS}$ distributions for class 8 suggest the presence of significant velocity gradients (up to 10 km s$^{-1}$) between the photosphere and the chromosphere. This is predominantly due to strong downflows in the chromosphere that become notably weaker in the photosphere. Large velocity gradients are evident in examples (c) and (e) of Fig.\ \ref{fig:8} at the locations of class 8 profiles and  likely contribute to the observed negative amplitude asymmetries.

The asymmetries in $V_R$ 7 and $V_R$ 8  were fitted in NLTE using two atmospheric components. For $V_R$ 7, two components with similar filling factors ($\mathit{ff}$) but different temperatures 
and LOS velocities were retrieved, see Fig.\ \ref{fig:15}. The cold component is non-magnetic and nearly at rest, while the hot component displays a velocity gradient with downflows in the lower layers and upflows higher up. The hot component exhibits a magnetic field that increases with height, starting near zero in the photosphere and reaching approximately 800 G at the height of the Mg line core maximum sensitivity,  
 where it becomes more horizontal. Such a magnetic configuration is consistent with the existence of a  canopy around bright magnetic structures. 

For $V_R$ 8, 
 the generally hotter component is slightly dominant, with $\mathit{ff}=0.57$, and resembles the stratification of Family A for bright magnetic elements. It displays a nearly constant velocity stratification with $v_{LOS}\sim 1.5$ km s$^{-1}$ and a magnetic field that reaches up to 500 G near the  Mg line core maximum sensitivity height. The second component is cooler in the lower and higher layers, with a weaker magnetic field, and shows a strong velocity gradient with a complex velocity profile, indicating downflows in the photosphere, upflows in the mid-layers, and downflows above $\log(\tau)\sim-4$.  The NLTE inversion results  support the scenario of 
both velocity and magnetic field gradients in a bright magnetic element as sources of large negative amplitude asymmetries in the Mg line Stokes $V$ profiles.
In these two cases of Family D, both the WFA and SIR results for the magnetic field strength are consistent with the weighted average of the two magnetic components.

 \begin{figure}[h!]
   \centering
             \includegraphics[width=0.75\hsize]{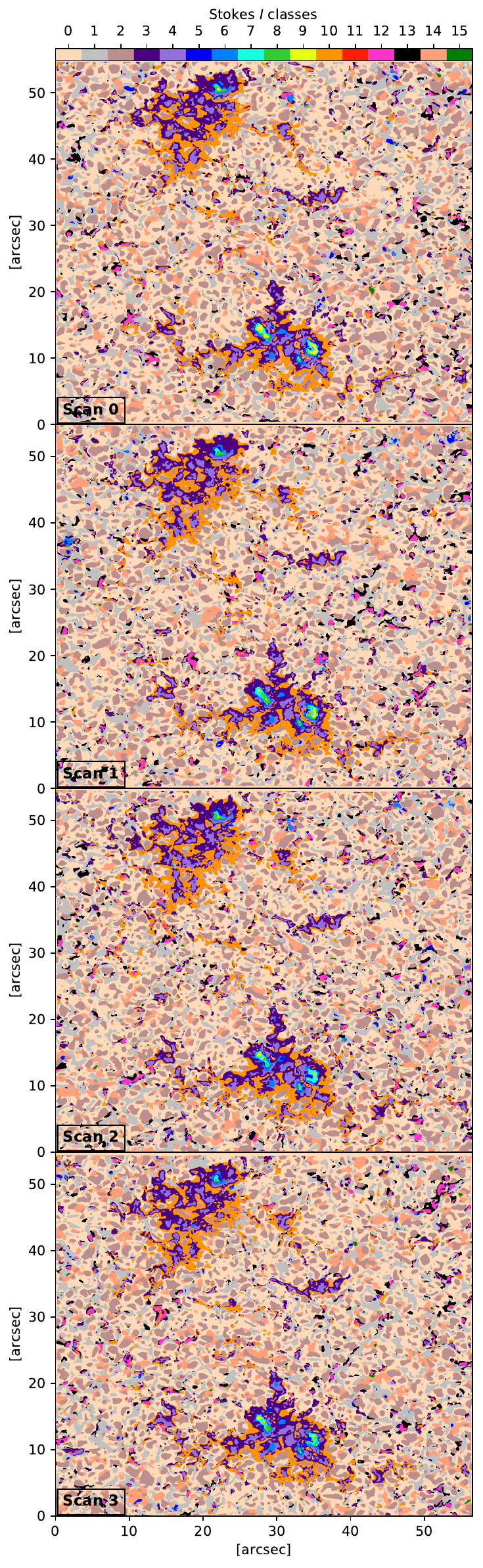} 
      \caption{Spatial distribution of the 
      16 classes of Stokes $I$ profiles of the Mg I b$_2$ line found in the four scans of our dataset, according to the $eud$ classification. The maps show the full FOV.
              }
         \label{fig:9}
   \end{figure}

 \begin{figure*}[!htbp]
   \centering
        
        \includegraphics[width=\hsize]{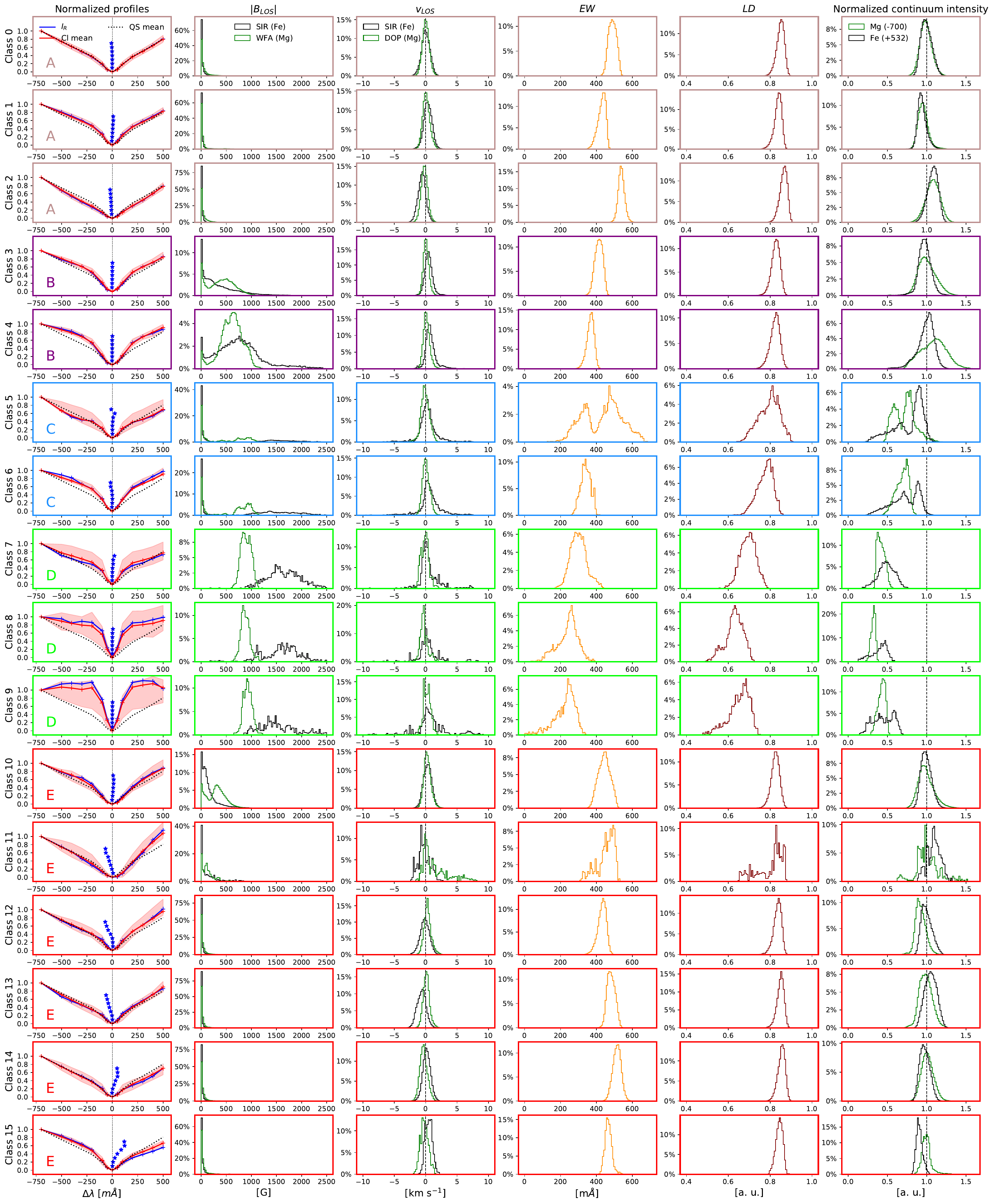}
      \caption{Spectral and physical properties of the 
      16 different classes of Stokes $I$ profiles. 
      From left to right: the $I_R$'s (blue) and the mean class profiles (red), their distributions for the unsigned longitudinal magnetic field, the LOS velocities, the equivalent width, the line depth, and  the normalized continuum intensities in the Mg line ($I_{-700}/I_{QS_{-700}}$) and in the Fe line ($I_{+532}/I_{QS_{+532}}$). Dotted profiles in the first column correspond to the mean QS, and the blue markers show the line bisectors of the $I_R$'s at different intensity levels. Similar format as in Fig.\ \ref{fig:7}. 
              }
         \label{fig:10}
   \end{figure*}

\section{Stokes $I$ classification} \label{sect:stokesI}

The maps in Figure \ref{fig:9} show the results of the $eud$ classification for the 
16 classes of Stokes $I$ profiles identified in our Mg dataset.
Similar to the Stokes $V$ case, most of the Stokes $I$ classes exhibit a consistent spatial distribution among the four different scans, 
except for the classes from 
11 to 15, which appear in different locations over time. These classes correspond to very asymmetric profiles that are likely associated with transient structures or rapid events. 
As with the Stokes $V$ classes, we grouped the Stokes $I$ classes  into different families based on their common characteristics.

\subsection{Family A}
Classes 0, 1, and 2 
are the largest groups in the Stokes $I$ classification 
($\sim39\%$, $17\%$, and $19\%$
of the total sample, respectively; see Table \ref{tab_res2}) and display a clear spatial correlation with the granular edges, intergranular lanes, and  bright granular centers, respectively. They are all fairly symmetric profiles, as indicated by their nearly centered line bisectors. However, mild deviations become noticeable at increasing intensity levels for all three classes (see blue markers in the first column of Figure  \ref{fig:10}). These deviations are slightly blueward in classes 0 and 2, reflecting the photospheric upflows present in the granules --- which are stronger in the granule center and nearly fade toward their edges --- while they are slightly redward in class 1 due to the photospheric downflows in the intergranular lanes. This is consistent with their histograms of the photospheric LOS velocity, which peaks nearly at zero for class 0, at positive values for class 1,  
 and at negative values for class 2; 
 see also Table \ref{tab_res2}. The Mg line-core Doppler velocities follow a similar pattern but with reduced contrast, indicating that the photospheric velocity pattern of the quiet Sun  remains present, though weaker, in the low chromosphere.
 
The spectral characteristics and associated physical quantities in the histograms of Fig.\ \ref{fig:10} and in Table \ref{tab_res2} present rather subtle differences between these classes, so we group them together and classify them as Family A. The main morphological difference 
is the equivalent width which covers smaller values in class 1 than in class 0, whilst it is slightly larger in class 2, consistent with their representative profiles $I_R$ and with the expected different temperatures of the associated structures. 
 Something similar happens with their normalized continuum intensity distributions, whose average values are in both Mg and Fe lines near 
$1$ for class 0, 
but are slightly lower for class 1 and slightly higher for class 2.
Likewise, classes 0 and 2 have slightly deeper profiles 
than those in class 1. 
The $|B_{LOS}|$ 
distributions for these three classes display a single peak near zero, which indicates that the LOS field strength is generally weak both in the photosphere and the chromosphere 
as expected in most quiet Sun areas. 
The average longitudinal magnetic field strength in the photosphere is 
18 G for class 0, 19 G for class 1, and  14 G for class 2. In the chromosphere, the average values are 
44 G, 39 G, and 53 G, respectively. However, in the quiet Sun, polarization signals are mostly well below the noise level in both spectral lines, rendering these magnetic field estimates less reliable and potentially influenced by noise. 
Some of these characteristics can also be seen in Figure \ref{fig:11}, which shows in examples (a) and (b)  a zoomed-in portion of the quiet Sun granulation and highlights the location of classes 1 (intergranules) and 2 (granule centers), respectively.

Figure \ref{fig:16} presents the results of the NLTE inversions  
for the representative profiles $I_R$'s in this family. 
As indicated in Table \ref{tab2}, profiles in Family A required a relatively low number of free parameters to achieve a good fit. The resulting stratification of the physical parameters is consistent with the typical quiet Sun structure in the photosphere:  upflows and temperatures slightly higher than the FALC model in the low photosphere for $I_R$ 2 (granule center), weaker upflows and temperatures  similar to FALC for $I_R$ 0 (granule border), and downflows with temperatures slightly cooler than FALC for $I_R$ 1 (intergranule). 
The temperature difference between $I_R$ 1 and $I_R$ 2 is nearly 200 K at $\log(\tau)=-0.7$.
In the mid-to-upper photosphere, above $\log(\tau)=-1.4$, we observe an inverse granulation pattern in the temperature, with the granule centers becoming about 100 K cooler than the intergranules. 
The temperature difference between $I_R$ 1 and $I_R$ 2 increases higher up, and becomes of the order of 500 K at $\log(\tau)=-4.4$.
For all three cases, the inversions retrieve weak and rather horizontal magnetic fields, in agreement with the WFA and SIR inversions. However, these values are likely not meaningful, given that the polarization signals in both lines are extremely weak and essentially at the noise level.

\begin{table*}[ht!]
\centering
\caption{Summary Table of Fig.\ \ref{fig:10}. }
\begin{tabular}{ c c c c c c c c c c c    } 
\hline
$F$ & Class  & $\%$ pix & mean  &  mean & mean  &  mean  & mean  & mean & mean & mean  \\
&  &   & ($|B_{LOS}^{SIR}|$) & ($|B_{LOS}^{WFA}|$) &($v_{LOS}^{SIR}$) & ($v_{LOS}^{DOP}$) & ($EW$) & ($LD$) & ($I_{c}/I_{QS}$) & ($I_{c}/I_{QS}$)\\
 &  &   & [G] & [G] & [km s$^{-1}$] & [km s$^{-1}$] & [m{\AA}] & & [Fe, +532] & [Mg, -700]\\
\hline
\multirow{3}{*}{A} & 0 & 38.78 & 18 & 44 & -0.01   & 0.01 & 492 & 0.85 & 1.00 & 1.00 \\ 
 & 1  & 16.79 & 19 & 39 & 0.36   & 0.10 & 451 & 0.84 & 0.94 & 0.91 \\ 
& 2  & 18.58 & 14 & 53 & -0.59   & -0.20 & 539 & 0.86 & 1.08 & 1.08 \\ 
\hline
\multirow{2}{*}{B} & 3  & 5.89 & 372/55/545 & 408/53/500 & 0.47   & 0.07 & 413 & 0.83 & 0.97 & 1.01 \\ 
& 4  & 3.16 & 760/58/821 & 604/54/629 & 0.75   & 0.16 & 366 & 0.82 & 0.99 & 1.09 \\ 
\hline
\multirow{2}{*}{C} & 5  & 0.30 & 743/26/1724 & 435/38/797 & 0.31   & -0.07 & 434/315/506 & 0.79 & 0.77/0.56/0.91 & 0.72/0.59/0.81 \\
& 6  & 0.30 & 1071/25/1639 & 591/34/846 & 0.65   & 0.06 & 340 & 0.78 & 0.72/0.61/0.90 & 0.69 \\
\hline 
\multirow{3}{*}{D} & 7  & 0.11 & 1730 & 894 & 0.53   & -0.12 & 300 & 0.69 & 0.49 & 0.43 \\ 
& 8  & 0.09 & 1754 & 880 & 0.37   & -0.26 & 240 & 0.65  & 0.45 & 0.42 \\ 
& 9  & 0.04 & 2017 & 926 & 0.85   & 0.21 & 223 & 0.63 & 0.39 & 0.32 \\ 
\hline 
\multirow{6}{*}{E}& 10  & 7.14 & 164 & 303/108/413 & 0.30   & 0.29 & 444 & 0.82 & 0.97 & 0.99 \\ 
& 11  & 0.03 & 125 & 130 & -0.26   & 1.47 & 452 & 0.80 & 1.09 & 1.01\\ 
 & 12  & 0.97 & 19 & 42 & -0.14   & 0.33 & 431 & 0.83 & 1.00 & 0.92  \\ 
& 13 & 1.47 & 12 & 31 & -0.65   & 0.12 & 482 & 0.85 & 1.05 & 0.99 \\ 
& 14  & 6.27 & 18 & 44 & 0.16   & -0.31 & 514 & 0.85 & 0.98& 1.01 \\
& 15  & 0.08 & 39 & 51 & 0.48   & -0.20 & 468 & 0.84 & 0.91& 1.00 \\ 

\hline 
\end{tabular}
\label{tab_res2}
\tablefoot{From left to right, columns indicate the family $F$, the class, the percentage of pixels in each class relative to the total pixel sample (recalling that all the Stokes $I$ profiles in the dataset were classified), and the mean values in the histograms of Fig.\ \ref{fig:10} for each class. For the classes clearly exhibiting two distinct populations in the histograms, three mean values are provided: one for the entire distribution and one for each population.}
\end{table*}

\subsection{Family B}
Family B is formed by classes 3 and 4. 
Both classes consist of narrow profiles with bright wings 
which are found over the bright magnetic structures, either in the plage regions around the pores or in isolated bright magnetic elements. These regions are referred to as bright because of their appearance in the Mg continuum intensity, but they are not consistently seen bright in the continuum of the Fe line.
 Class 
 4 profiles tend to appear in the core of these structures and are surrounded by class 
3 profiles which concentrate 
 around the magnetic core of the bright structures where the continuum intensities are slightly lower, such as in examples (g) to (j) of Fig.\ \ref{fig:11}. 
This explains why the photospheric $|B_{LOS}|$  and the continuum intensity distributions for class 
4 in Fig.\ \ref{fig:10} reach higher values than those for class 
3. Also, class 
4 profiles generally display brighter wings, which results in smaller equivalent widths than class 
3 profiles 
 and are associated with faster downflows in the photosphere. 

Family B of Stokes $I$  is morphologically correlated with family A of Stokes $V$. The slight differences in wing brightness between Stokes $I$ classes 3 and 4 explain, within the weak field regime, the mild variations in lobe width observed between Stokes $V$ classes 0 and 1. Although the spatial correlation between these classes is not strictly one-to-one, both families are consistently observed at the core of bright magnetic regions.

The NLTE inversions of  
$I_R$ 4 profiles retrieve hot atmospheres and magnetic fields of several hundred Gauss, with inclinations that are 
largely oblique (see Fig.\ \ref{fig:16}). 
These results are expected, considering that   
$I_R$ 4 belongs to a bright magnetic element. However, the temperatures around $\log(\tau)=0$ are nearly the same as the FALC model.  
This is consistent with these structures displaying  continuum intensities in the Fe line comparable to that of granules, while the excess brightness  observed in the Mg pseudo-continuum at $-700$ m\AA\ indicates that these magnetic structures start being hotter than their surroundings slightly higher in the atmosphere, above $\log(\tau)=-0.5$.
The resulting LOS velocity and magnetic field strength from the classical methods fall within the errors of the inversions and are consistent with the NLTE stratifications.

\subsection{Family C}
\begin{figure*}[htp!]
   \centering
            
             \includegraphics[width=\hsize]{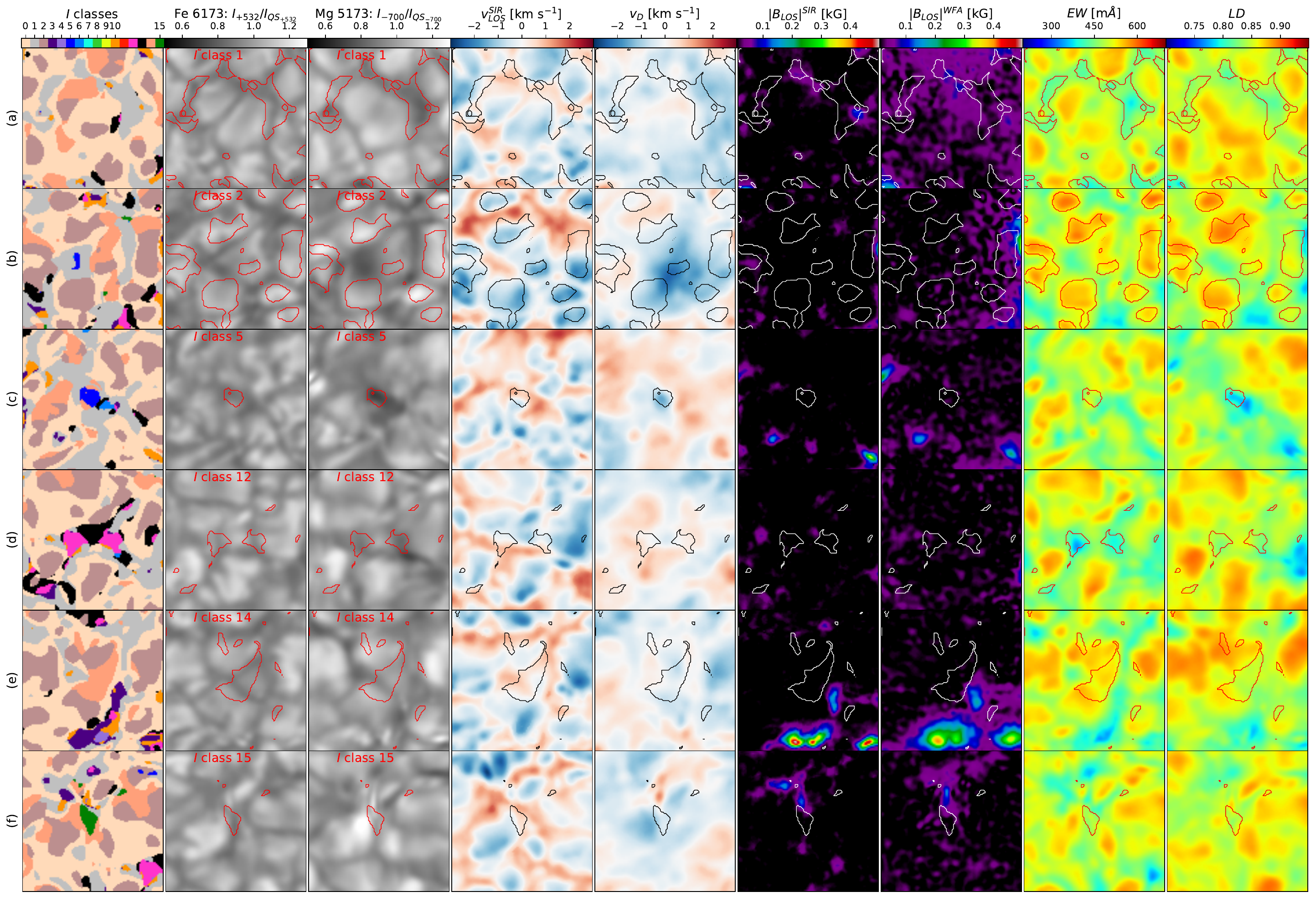} 
        \includegraphics[width=\hsize]{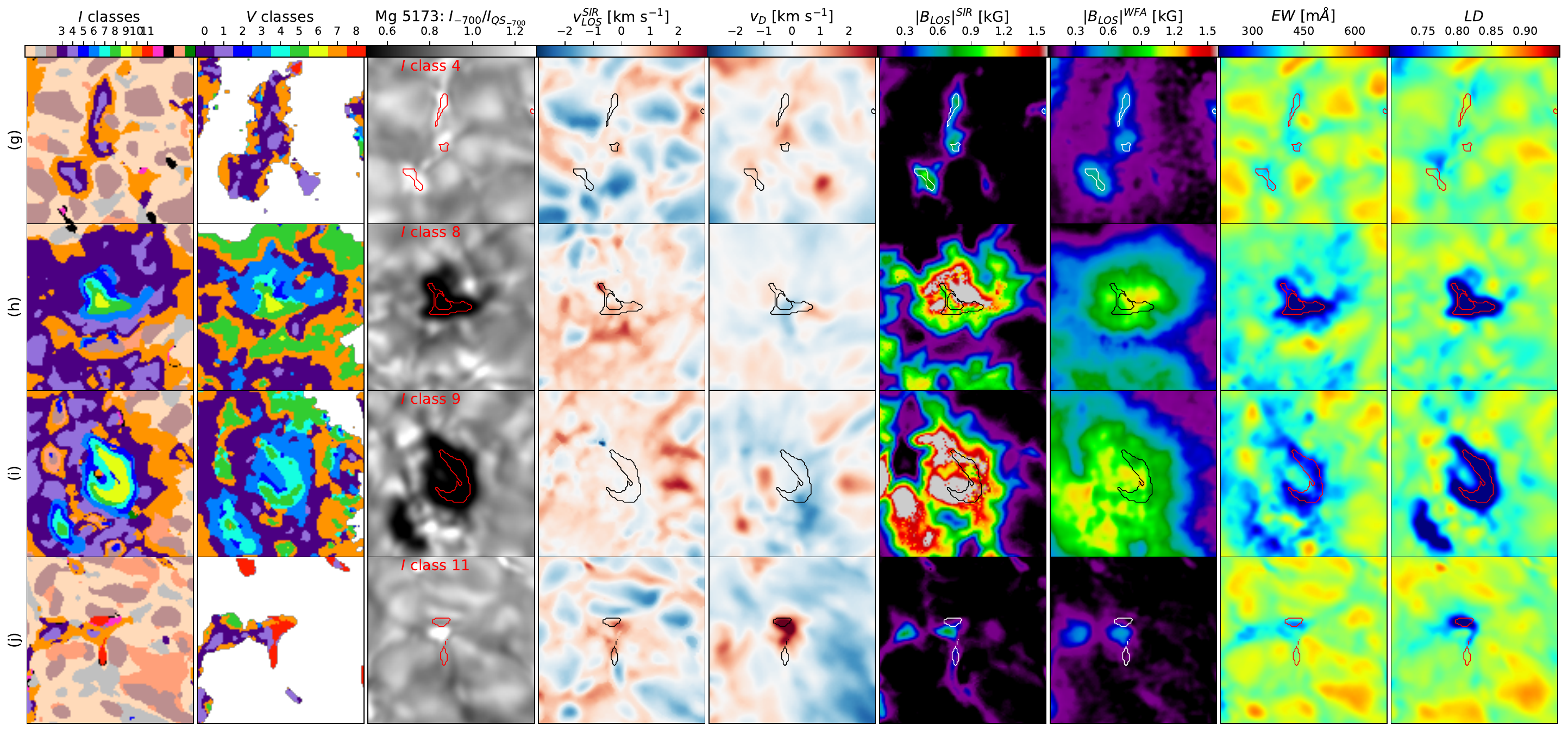}
               \caption{Examples of different structures and their associated Stokes $I$ classes and physical properties.  For examples (a)-(f) panels show, from left to right: the $eud$ classification of Mg Stokes $I$,  Fe continuum intensity (at +532 m{\AA}) normalized to its mean QS value, Mg continuum intensity (at -700 m{\AA}) normalized to its mean QS value, photospheric LOS velocity from the SIR inversions of the Fe line, Doppler velocity of the Mg line core, unsigned LOS magnetic field from the SIR inversions and the WFA in the Mg $\pm100$ m\AA\ spectral window, the  equivalent width, and the line depth of the Mg Stokes $I$ profiles. All the examples show a $\sim 6\arcsec \times 6\arcsec$ sub-field. For each row, the contours enclose the Stokes $I$ class specified in the intensity maps.  Columns in examples (g)-(j) are similar to those in examples (a)-(f), except for the second column which displays the associated Stokes $V$ classes in the magnetic structures. Also, examples (a)-(f) use a  smaller magnetic field scale than examples (g)-(j), allowing for the visualization of weaker magnetic field regions.
              }

         \label{fig:11}
   \end{figure*}

Family C is formed by 
classes 5 and 6, which appear associated with both magnetic and non-magnetic regions, as portrayed by the $|B_{LOS}|$ histograms in Fig.\ \ref{fig:10}.
This is similar to Family B classes and, as in the case of most  Stokes $V$ classes, indicates that some of the intensity profiles can display similar morphology 
across different physical scenarios.  
 However, class 5 exhibits a binomial distribution of $EW$, with a population peaking at 
 $EW \sim 315$ m\AA\ and a second peak near 
 $500$ m{\AA} (see Fig.\ \ref{fig:10} and Table \ref{tab_res2}). 
 This could be attributed to the $eud$ method's inability to separate the different features of $I_R$ 5, which exhibits both a narrow line core and wider wings --- a peculiar morphology that may result from emissions in the inner wings of the line or enhanced absorption in the outer wings. 
  Consequently, this class contains both narrow and wide intensity profiles. 
 The narrower profiles in class 5 are primarily located at the outer boundary of pores (see examples (h) and (i) in Fig.\ \ref{fig:11}), and are thus associated with the stronger field population in the $|B_{LOS}|$ histogram and with the lower continuum intensity populations. On the other hand, the wider profiles in class 5 are situated in quiet Sun regions, specifically within certain intergranular lanes that appear notably dark in the Mg continuum image, as shown by example (c) in Fig.\ \ref{fig:11}, although they are still brighter than the pores. This second population is thus associated with the weaker fields and with the relatively brighter profiles of this class. Nonetheless, the majority of profiles in class 5 exhibit a morphology similar to that of the representative profile, as reflected in the average class profile (see red profiles in Fig.\ \ref{fig:10}).

 Although not a perfect match, there is some spatial correlation between this family and family B of Stokes $V$ at the periphery of the pores. In particular, Stokes $I$ classes 5 and 6 correspond to Stokes $V$ classes 2 and 3, respectively. This alignment is expected since, in the weak-field regime Stokes $V$ is proportional to the derivative of Stokes $I$. Therefore, the different wing shapes observed in Stokes $I$ classes 5 and 6 would naturally result in diverging and converging Stokes $V$ wing signals, as seen in Stokes $V$ classes 2 and 3, respectively. However, in the pore regions the Mg line may not be entirely in the weak-field regime (see, e.g., Paper I), meaning the one-to-one correspondence between Stokes $I$ and Stokes $V$ is not always strictly satisfied.

\begin{figure*}[ht!]
   \centering
         
         \includegraphics[width=\hsize]{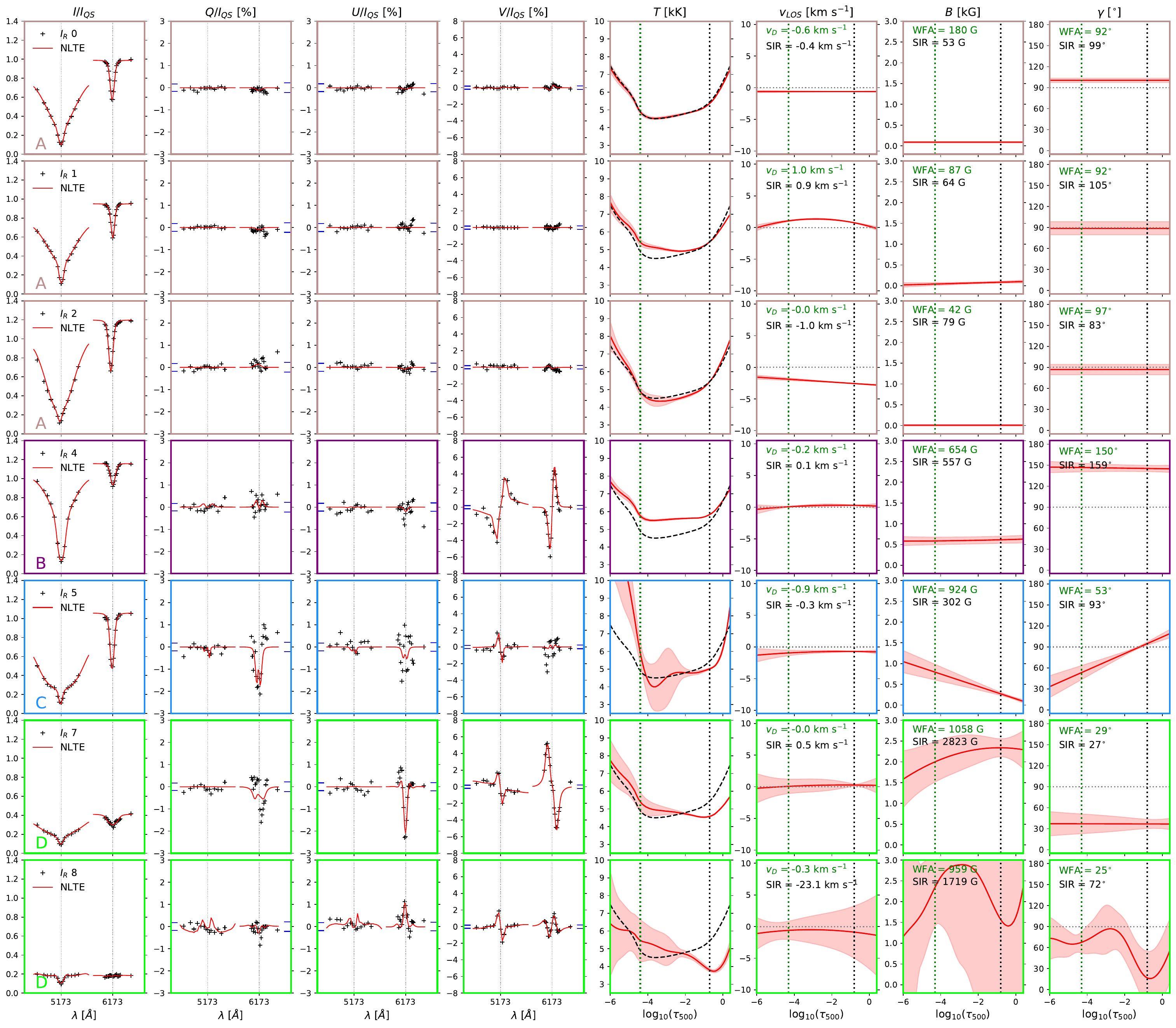} 
      \caption{NLTE inversions of selected symmetric Stokes $I$ representatives as indicated in the legends on the first column. The family is indicated in the left bottom corner of the panels in the first column. 
      Same format as in Fig.\ \ref{fig:15}.              }
         \label{fig:16}
   \end{figure*}

The representative $I_R$ 5 belongs to the periphery of a pore, and the NLTE inversions  in Fig.\ \ref{fig:16} recover an atmosphere that displays a complex temperature profile which rapidly cools down in the low photosphere but slightly rises again above $\log(\tau)\sim-2$ and shows a second decrease at $\log(\tau)\sim-3$ (fluctuations that might not be significant due to the large uncertainties estimated at these heights) before steeply rising in the higher layers, where it reaches values larger than ten thousand Kelvin. 
 The retrieved atmosphere displays a magnetic field that increases with height from about 300 G at $\log(\tau)=0$ to $\sim 800$ G at $\log(\tau)=-4.3$, in agreement with the WFA and the SIR inversions. These results are consistent with the formation of a magnetic canopy around pores.

 \begin{figure*}[ht!]
   \centering
        
          \includegraphics[width=\hsize]{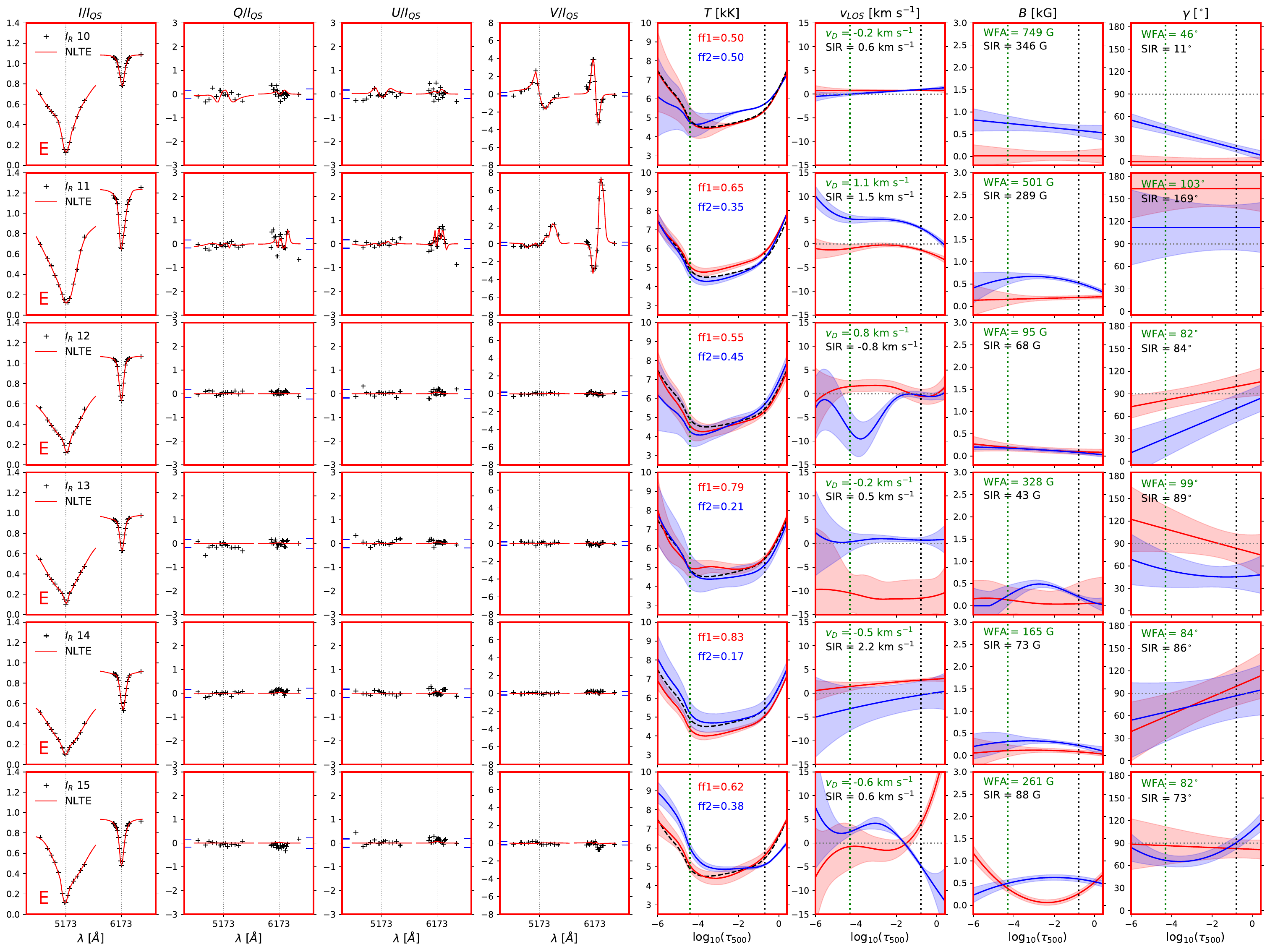} 
      \caption{NLTE inversions of asymmetric Stokes $I$ representatives. Classes from 
      10 to 15 are displayed from top to bottom (see legends in the first column). Same format as in Fig.\ \ref{fig:15}. 
              }
         \label{fig:17}
   \end{figure*}

\subsection{Family D}
Family D is formed by classes 
7, 8, and 9 of Stokes $I$ profiles, which are all located within the pores, as illustrated in examples (h) and (i) of Fig.\ \ref{fig:11}. 
Unlike 
family C, these classes are exclusively found inside the pores, which explains why their $|B_{LOS}|$ distributions only display one population of kiloGauss magnetic fields. 
Family D contains the classes with
 the narrowest profiles in the sample (with average $EW$ of 
 300, 240, and 223 m{\AA}, respectively) and also the shallowest  (with average $LD$ of 
 0.69, 0,65, and 0.63, respectively). They also exhibit the lowest continuum intensities in both spectral lines, 
 along with a significant degree of emission in their wings. 
As indicated by the histograms of Fig.\ \ref{fig:10}, the estimated LOS velocities for some profiles in these classes are notably high, particularly in the photosphere. However, such estimations are not reliable due to the rather flat intensity profiles that both spectral lines typically exhibit in these classes, especially in the Fe line  (see, for example, $I_R$ 
8 in Fig.\ \ref{fig:16}). Such flat profiles significantly affect the performance of the SIR inversions to properly fit the Fe line.

The Stokes $I$ classes 7, 8, and 9 in this family spatially and morphologically correlate with Stokes $V$ classes 4, 5, and 6 in family C, which are all found inside the pores. This is consistent with the WFA, which predicts nearly constant non-zero Stokes $V$ wing signals (as seen in Stokes $V$ class 4) for intensity profiles with a constant wing slope (Stokes $I$ class 7), nearly zero Stokes $V$ wing signals (Stokes $V$ class 5) for intensity profiles displaying  wing emissions close to the continuum level (Stokes $I$ class 8), and opposite polarity Stokes $V$ signals (Stokes $V$ class 6) for strong wing emissions exceeding the continuum intensity (Stokes $I$ class 9). However, due to the strong magnetic fields inside the pores, the WFA is not always strictly applicable to the Mg line. As a result, the correlation between individual classes is not always one-to-one but is better preserved at the family level (see examples (h) and (i) in Fig.\ \ref{fig:11}).

The NLTE inversions of $I_R$ 7 and $I_R$ 8 in Fig.\ \ref{fig:16}, which belong to the interior of a pore, retrieve very low temperatures in the lower layers of the atmosphere (down to approximately $4500$ K for $I_R$ 
7 and around $3700$ K for $I_R$ 
8) and temperatures hotter than the FALC model above $\log(\tau)\sim-2$, where the estimated errors become considerably large partly due to the large number of nodes used for the temperature in order to reproduce the observed profiles. 
The resulting magnetic field decreases with height and is largely vertical for $I_R$ 
7, as expected in pores and in agreement with the LTE methods, although there is a discrepancy of various hundred Gauss between the methods. 
In contrast, there is not good agreement between the NLTE and the classical inferences of $v_{LOS}$, $B$, and $\gamma$ for $I_R$ 
8. Despite the NLTE inversions succeeded in reproducing the observed profiles of  $I_R$
8, the estimated uncertainties are considerably large even in the lower layers, which means that the NLTE results could have a large error for these parameters at all heights. 
 However, the temperature below $\log(\tau)=-2$ seems to be better determined and displays the lowest temperatures among all the inverted $I_R$'s, which is consistent with the very low intensity levels measured in both spectral lines that emerge from the core of a pore.
Such overly low intensities and very shallow observed profiles, in combination with the presence of some emission in the wings of the Mg line, 
also impact the performance of the classical methods used to infer the LOS velocities and the magnetic field in the photosphere and in the chromosphere. In particular, the SIR inversions deliver unusually large photospheric LOS velocities of $-23$ km s$^{-1}$, values that are not reliable since such inversions failed to produce a reasonable fit in this pixel (not shown in Fig.\ \ref{fig:16}).

\subsection{Family E}
We group all asymmetric profiles in classes  
10 to 15 into Family E.
Class 10 profiles display asymmetric wing brightness and appear agglomerated at the borders of the bright magnetic structures. They spatially correlate with the asymmetric Stokes $V$ profiles in class 7 and display similar physical properties to Stokes $I$ family B found within the bright structures, although exhibiting slightly darker intensities and magnetic fields that increase from the photosphere to the chromosphere (see Table \ref{tab_res2}).
In particular, the blue-wing asymmetry of $I_R$ 10 was reproduced using two atmospheric components with the same $\mathit{ff}$ (see Figure \ref{fig:17}). One component  is nearly non-magnetic and displays a temperature profile similar to the FALC model. The second component has a hotter temperature profile up to $\log(\tau)=-4.4$ and shows a magnetic field that increases with height from $\sim 500$ G in the photosphere to $\sim 750$ G at $\log(\tau)=-4.3$, becoming more horizontal. These results are consistent with the WFA and SIR inversions, considering the associated uncertainties. These atmospheres also reproduce the mild amplitude asymmetry of the Mg I b$_2$ Stokes $V$ profile (which corresponds to Stokes $V$ class 7). Again, for $I_R$ 10 profiles, the results support the existence of a low-lying magnetic canopy around the bright magnetic structures due to the opening of the magnetic field with height. This configuration can be modeled as two different atmospheric components along the LOS: component 1 (red model) represents the quiet Sun or stray light within the pixel, while component 2 (blue model) represents the magnetic canopy of the bright structure. Together, these components successfully reproduce the observed asymmetries in the Stokes $I$ and $V$ profiles.
   
  Class 11 profiles exhibit the largest blueward asymmetries and are associated with the largest downflows inferred for the low chromosphere, 
as reflected in the $v_{D}$ distributions of Fig.\ \ref{fig:10} and Table \ref{tab_res2} (average $v_{D}$ of $1.47$ km s$^{-1}$), while in the photosphere, class 12 profiles are mostly associated with upflows (average $v_{LOS}=-0.26$ km s$^{-1}$).
 This suggests the presence of large velocity gradients along the LOS. 
 These profiles are observed at the edges or within some bright magnetic structures, as shown in example (j) of Fig.\ \ref{fig:11}. The $|B_{LOS}|$ distributions display moderate values but do not expand beyond 1 kG.
This is the smallest group in the classification, and their associated Stokes $V$ profiles display very large negative asymmetries (i.e., they belong to class 8 in the Stokes $V$ classification). 
  
 Classes 
 12 and 13  both exhibit blueward asymmetries, as highlighted by the line bisectors (blue markers in Fig.\ \ref{fig:10}). However, histograms in Fig.\ref{fig:10} and Table \ref{tab_res2} show that class 
 13 profiles are slightly wider, brighter, and deeper profiles, with larger asymmetries than class 12 profiles.  
 The histograms reveal a preference for upflows in the photosphere in both classes, although they are faster in class 
 13 than in class 12.  
 In the chromosphere, 
 both classes show a preference for downflows, which are in average faster for class 12. Nonetheless, class 13 suggests larger velocity gradients between photosphere and chromosphere than class 12.
Both classes tend to appear together, as seen in example (d) of Fig.\ \ref{fig:11}, which depicts a small granule surrounded by dark integranular lanes, with class 
12 located  toward the edges of such granule and class 
13 in the center.
In this example, continuum intensities and $EW$ maps display characteristics typical of intergranules at the location of class 
12, but the $v_{LOS}$ maps indicate weak photospheric upflows. These structures, due to their small size and slightly brighter appearance in the photosphere than in the chromosphere, are likely newly emerging granules. This scenario is expected to generate velocity gradients with height and possibly multiple atmospheric components with different temperatures, thus explaining the profile asymmetries and their varying spatial distribution across different scans.

Classes 
14 and 15 contain profiles with large redward asymmetries, with class
15 profiles exhibiting a more extreme asymmetry and being less abundant compared to class 14. 
The different degrees of asymmetry are illustrated by the bisectors of the representative profiles in Fig.\ \ref{fig:10} and the average profiles in each class. The $v_{LOS}$ distributions in both classes indicate a preference for downflows in the photosphere,  generally faster in class 11, 
 and upflows in the higher layers 
 with larger velocity gradients in class 
 15 than in class 14.
Example (e) of Fig.\ \ref{fig:11} shows class 14  profiles located above the junctions of two or more granules with weak upflows in the chromosphere and weak downflows in the photosphere. 
 The faint dark lanes observed in the Mg continuum intensity at the junctions of these granules are not well defined in the maps of equivalent width and line depth. These structures might correspond with granules undergoing splitting or the formation of a new intergranular lane.  Such processes likely occur over shorter periods than the temporal cadence of our dataset, which could explain the different locations of class 14 throughout the four scans. The splitting of granules likely occurs first at photospheric levels, which might explain the lower continuum intensities in Fe than in Mg.
Class 15 profiles are usually surrounded by class 14 profiles, also at the junctions of various granules, as in example (f) of  Fig.\ \ref{fig:11}. The larger redward asymmetries of their wings suggest stronger photospheric downflows and likely correspond to a later stage of the splitting process, which could also explain the lower continuum intensity distributions for this class.

Figure \ref{fig:17} presents the NLTE results for the representatives $I_R$'s in Family E. 
 These profiles are located in complex sites,  and the cadence of our observations might not be sufficient to fully understand their nature, as their origin might be associated with faster events. Notwithstanding, the asymmetries were successfully reproduced by using two atmospheric components, each with initially distinct LOS velocities. Furthermore, all physical parameters in both atmospheres were allowed to vary, as indicated in Table \ref{tab2}, thus requiring a substantial number of free parameters. The resultant  two-component atmospheres show different and intricate gradients with height in all cases.
Additionally, the estimated errors for this family are somewhat large across the entire height range of the atmospheres, 
making the interpretation of the results and the comparison with the classical methods not straightforward.

%
\section{Discussion and conclusions}

Although various classifications of Stokes $V$ profiles exist in the literature for photospheric spectral lines \citep[e.g.,][]{Sigwarth1999, Sigwarth2001, Khomenko2003, Dominguez2003, Viticchie2011, Bellot2019}, the diversity of polarization shapes in chromospheric spectral lines has been studied to a lesser degree, often using semi-empirical approaches  \citep[e.g.,][]{Quintero2018, Quintero2019, Carlin2019}. 
In this study, we performed a detailed morphological classification of the Stokes $I$ and $V$ profiles of the Mg I b$_2$ line at 5173 {\AA}, using high-resolution spectropolarimetric observations of a portion of the solar atmosphere containing a bipolar magnetic region (pores and plage), which was surrounded by quiet Sun granulation. Our classification, based on the careful selection of a small number of representative profiles and the Euclidean distance method, has provided insights into various solar magnetic structures, their physical properties, and the processes occurring at different heights in the solar atmosphere.

The Euclidean distance method proved to be an effective approach for classifying Mg I b$_2$ profiles, as evidenced by the strong resemblance between the average profiles in each class and their respective representative profiles. This ensures that the majority of the profiles were morphologically classified correctly. However, our approach required an extensive visual inspection of the dataset and the manual selection of the representative profiles --- a task that is both demanding and time-consuming. Nonetheless, this step was necessary, as our primary goal was to identify all distinct morphological variations present in the observed solar scenery, regardless of their Doppler shifts and amplitudes.
Unsupervised methods, such as k-means clustering, offer an alternative approach and have been widely applied in solar studies \citep[e.g.,][]{Sainz2019, Kuckein2020, Moe2023}. These methods are computationally efficient and robust, but their performance depends on selecting an optimal number of clusters ($k$), which must be specified as an input parameter. Moreover, they may struggle to distinguish subtle yet important differences between profiles. While these limitations can lead to an underestimation or overestimation of the number of classes --- which is crucial for the purposes of this study --- unsupervised methods remain valuable for other purposes, such as identifying dominant features in the dataset or to speed up inversions.

Understanding the variety of shapes and polarization signatures of
the Mg I b$_2$ line at 5173 \AA\ provides valuable information about the physics of the solar atmosphere and the magnetic field configuration in the region where the line is formed, namely,  the layers between the photosphere and the chromosphere comprising the so-called "temperature minimum" region. Since the line forms under NLTE conditions \citep[e.g.,][]{Lites1988, Quintero2018}, a proper interpretation of the underlying physics, with emphasis on the thermodynamic properties, requires a formal NLTE approach.

Nonetheless, in this work, some of the physical properties, such as the longitudinal magnetic field and the LOS velocities, were inferred using classical methods on the Mg line. This choice was driven by the significant computational time required for the NLTE spectropolarimetric inversions to process the entire dataset, which consisted of nearly $4\times10^{6}$ profiles.
Instead, we applied the WFA \citep[e.g.,][]{Deglinnocenti2004} to the Mg line core to infer the magnetic field configuration, and the Doppler shift of the Mg line core to estimate LOS velocities in the low chromosphere. Additionally, SIR inversions \citep{Ruiz1992} of the Fe line in LTE were performed to obtain complementary information about the photosphere.

 The validity of these methods on the Mg line, despite its NLTE nature, was addressed in Paper I, where we found that both the WFA and the Doppler shift method  provide qualitatively reliable results that can be used as a statistical reference for the longitudinal field and the LOS velocities in the low chromosphere, respectively, while significantly reducing the computation time when dealing with large datasets \citep[e.g.,][]{Vukadinovic2022}.
 However, special care must be taken when dealing with  profiles that exhibit large asymmetries and peculiar shapes,  as the results from these classical methods may not be accurate, given that they do not account for height gradients within the line formation region.
 
Additionally, the WFA becomes unreliable in regions of strong magnetic fields, such as the magnetic cores of pores where the chromospheric $|B_{LOS}|$ can exceed 1400 G in a number of pixels. 
 According to \citet[][]{Vukadinovic2022},  the LOS magnetic field retrieved by the WFA in the Mg line remains reliable  up to 1400 G, even in the presence of moderate velocity gradients. In Paper I, we found that the results from the WFA and NLTE inversions disagree by less than 100 G in weak field regions of the low chromosphere, but the discrepancy increases to about 300-400 G for $|B_{LOS}|>1400$ G in the NLTE scheme, with the WFA generally estimating weaker fields than the NLTE inversions. However, 
 the NLTE inversions presented in Paper I 
 were performed with a simplified setup to reduce computational time. 
While this setup provided good fits for most of the FOV, the fits were notably poorer in the cores of the pores. 
 Additional sources of error in the NLTE results may include the use of a simplified atomic model and the spectral sampling of the observations, which, if insufficient, could directly affect the performance of the inversions, particularly in the core of the line.

We also performed NLTE inversions of 
selected representative profiles
 using the full Stokes information from both spectral lines, Mg and Fe, with the DeSIRe inversion code \citep{Ruiz2022}, treating the Fe line in LTE and the Mg line fully in NLTE. 
The  inversion setups were individually tailored to the complexity of each representative profile, providing the code with sufficient degrees of freedom to closely reproduce the observed spectral and polarization features.
 This approach allowed us to achieve good-quality fits, generally yielding reliable information on the temperature stratification from the photosphere to the chromosphere, associated with the different profile morphologies. To our knowledge, this is the first time that the Mg I b$_2$ line at 5173 {\AA} has been inverted using a full NLTE treatment on high-resolution spectropolarimetric observations. 
 The results of these inversions agree with those from Paper I  and suggest that in strong field regions, such as in the magnetic core of the pores (see, for example, $I_R$ 
 7 and $I_R$ 8 in Fig.\ \ref{fig:16}), the WFA underestimates the longitudinal component of the magnetic field with respect to the NLTE scheme by a few hundred Gauss. However, the estimated uncertainties for the NLTE inversions are quite substantial in these cases, making it difficult to accurately determine the magnetic field configuration.

The selection of representative profiles was done manually, using as fewer classes as possible to capture the broad range of variation in the spectra of the Mg line.
We identified nine distinct classes of Stokes $V$ profiles within the magnetic regions and 
16 classes of Stokes $I$ profiles across the entire FOV of the observations. 
These classes were grouped into families, not only because they tend to appear clustered spatially and temporally, but also because they share morphological characteristics and similar physical properties.
 
In the magnetic regions, we identified nine Stokes $I$ classes that morphologically correspond to the nine Stokes $V$ classes, as expected in the weak-field regime, where Stokes $V$ profiles are proportional to the derivative of Stokes $I$ \citep[e.g.,][]{Deglinnocenti2004}. However, while a strong spatial correlation exists between these classes, it is not always strictly one-to-one at the pixel level, with the correspondence being more consistent at the family level in most cases. Several factors could contribute to this discrepancy, including the higher noise level in Stokes $V$, which influences its morphology and affects the classification, particularly at the edges of magnetic structures where the signal-to-noise ratio approaches the   5$\sigma$ threshold value. Additionally, as discussed in Paper I, the Mg I b$_2$ line may not fully satisfy the weak-field approximation in the pores, leading to deviations from the expected proportionality between Stokes $V$ and the derivative of Stokes $I$ in strong-field regimes. 

 Thus, we described four families of Stokes $V$ classes (A-D) and five families of Stokes $I$ classes (A-E), each primarily associated with specific solar features as follows:

\begin{enumerate}[i)]

  \item In the quiet Sun granulation (granules and intergranules), no Stokes $V$ classification was performed due to the very weak signals detected in these regions, which fall well below the 5$\sigma$ threshold value selected for the classification.
    This is expected, given the typically weak magnetic fields present in the quiet Sun chromosphere \citep[e.g.,][]{Quintero2018, Bellot2019}. We found that the average longitudinal field strength in these areas is below 100 G, both in the photosphere and the chromosphere. However, on average,  it is approximately $2-4$ times stronger in the chromosphere. This could be due to the presence of different magnetic canopies that expand into the low chromosphere above the granules and intergranules, originating from  nearby photospheric magnetic field concentrations, such as isolated magnetic points embedded within the intergranular lanes \citep[e.g.,][]{Martinez2009} or from larger structures such as plage and pores. However, as the polarization signals measured in both spectral lines are generally at the noise level in the quiet Sun, the inferred magnetic field values in the photosphere and chromosphere are not robust.
     We also found that the intensity profiles are generally deep and largely symmetric in the quiet Sun areas (Family A of Stokes I), although the line bisectors show small deviations from the central wavelength  close to the continuum level as a result of the moderate LOS velocities associated with the photospheric convective flows, similar to what is typically observed in most photospheric spectral lines \citep[e.g.,][]{Bellot2006, Gonzalez2020}. The NLTE inversions of the representative profiles return larger temperature differences between granules and intergranules in the low chromosphere than in the photosphere, and the temperature reversal occurs at $\log(\tau)=-1.4$, which is consistent with previous observations and simulations of the reverse granulation pattern of the upper photosphere \citep[e.g.,][]{Ruiz1996, Borrero2002, Cheung2007}. We also found largely asymmetric Stokes $I$ profiles in different sites of the quiet Sun (classes 12 to 15 contained in Family E), which were reproduced using two-component NLTE inversions that display strong and complex velocity gradients. These profiles are rare ($\sim 9 \%$ of the total sample) and cannot clearly be associated with specific structures. To understand the nature of these profiles we need higher-cadence observations  with better spatial and spectral resolutions. Therefore, this might be a suitable study to be addressed with DKIST \citep{Rimmele2020} and EST \citep{Quintero2022}.
     
 \item The bright magnetic structures, as seen in the Mg pseudo-continuum, include bright isolated magnetic points and plage that are generally not seen as bright structures in the low photosphere. We found that the Mg line displays pretty symmetric Stokes $V$  (Family A) and Stokes $I$ profiles (Family B)  in their magnetic cores, which are associated with several hundred Gauss to nearly kiloGauss
  photospheric fields that either weaken with height 
  or remain nearly constant in the low chromosphere. The intensity profiles are narrow due to enhanced brightness in the wings which, according to the NLTE inversions, indicate enhanced temperatures in the upper layers of the atmosphere, but quiet Sun temperatures in the low photosphere. Further out, near the edges of these magnetic structures we found larger asymmetries in both Stokes $V$ (class 7 in Family D) and $I$ profiles (class 
  10 in Family 
  E), which were reproduced with two atmospheric components of different temperatures and magnetic fields that are consistent with the presence of low-lying magnetic canopies, where the magnetic field opens rapidly with height producing strong gradients along the LOS. This agrees with the work of \citet[][]{Kuckein2019}, who found similar canopy effects and asymmetries in the  Stokes $I$ and $V$ profiles of the Si I 10827 \AA\ line around magnetic bright points that were reproduced with two-component inversions.
 Another type of asymmetric Stokes $V$ profiles (with negative amplitude asymmetries --- class 8 in Family D) were also observed in the vicinity of bright magnetic structures and were associated with largely asymmetric Stokes $I$ profiles (class 
 11 in Family E). These asymmetries all portray strong chromospheric downflows producing large gradients along the LOS, and were reproduced by the NLTE inversions  with two components displaying complex velocity profiles. 
 
 \item The periphery and the interior of pores display narrow profiles in both Stokes $V$ (Families B and C) 
 and Stokes $I$ (Families C and D), associated with kiloGauss longitudinal fields in the photosphere, which decrease with height to nearly kiloGauss strengths in the low chromosphere. The intensity profiles become gradually shallower and flatter as the magnetic core of the pores is approached, indicating progressively cooler photospheric temperatures up to the order of $4000-4500$ K in agreement with previous studies \citep[e.g.,][]{Sobotka2003, Thomas2004, Sobotka2012}. Moreover,  the wings display some emission in the core of the pores. This characteristic is reproduced by the NLTE inversions with an increase in the temperature profile in the higher layers of the atmosphere, with temperatures that exceed 5000 K in the low chromosphere. 
 The Stokes $V$ profiles also display different signatures and signal levels in the wings, generally decreasing as approaching the magnetic core of the pores and even changing signs in the very core of the pores as a consequence of the prominent emissions in the wings of the Stokes $I$ profiles observed on these sites (see, e.g., set of profiles for $V_R$ 6 in Fig.\ \ref{fig:15}). 
  As explained by \citet[][]{Carlin2019}, different wing signals can also be generated in the presence of strong magnetic fields, NLTE effects, and atomic polarization, so that it is possible that the different classes of Stokes $V$ profiles observed in the pores are due to dichroism and magneto-optical effects.

\end{enumerate}

We aim to extend this analysis by including observations of the Mg I b$_2$ line in other solar structures such as sunspot umbrae, penumbral filaments, and light bridges, among others. This will allow us to provide a more complete morphological classification of the intensity and circular polarization profiles of this line and to better understand the physics of the temperature minimum region of the solar atmosphere. A full characterization must also include the linear polarization profiles, Stokes $Q$ and $U$, to account for the horizontal configuration of the magnetic field in the low chromosphere. These studies will soon be addressed with data from the recently completed SUNRISE III mission \citep[e.g.,][]{Barthol2011}, DKIST \citep{Rimmele2020}, and the upcoming high-resolution observations of EST \citep{Quintero2022}.

\begin{acknowledgements}
      We would like to thank the anonymous referee for important suggestions that greatly improved the content of this work. We also thank
       Edgar Carlin for valuable discussions and useful suggestions on this work and Mats G. L\"ofdahl for his assistance during the data reduction. A.S.T. has been funded by Consejer\'ia de Transformaci\'on Econ\'omica, Industria, Conocimiento y Universidades from Junta de Andaluc\'ia through grant POSTDOC-21-00832. This work was supported by the Spanish Ministry of Science and Innovation through the project PID2021-125325OB-C51. The authors acknowledge financial support from the Severo Ochoa grant CEX2021-001131-S funded by MICIU/AEI/10.13039/501100011033.  R.G. acknowledges the support by Funda\c{c}\~{a}o para a Ci\^encia e a Tecnologia (FCT) through the research grants UIDB/04434/2020 and UIDP/04434/2020. The Swedish 1 m Solar Telescope is operated on the island of La Palma by the Institute for Solar Physics of Stockholm University in the Spanish Observatory del Roque de los Muchachos of the Instituto de Astrofísica de Canarias. The Institute for Solar Physics is supported by a grant for research infrastructures of national importance from the Swedish Research Council (registration number 2017-00625).

\end{acknowledgements}

\bibliographystyle{aa}
\bibliography{Mgclass.bib}

\begin{thebibliography}{54}
\expandafter\ifx\csname natexlab\endcsname\relax\def\natexlab#1{#1}\fi

\bibitem[{{Barthol} {et~al.}(2011){Barthol}, {Gandorfer}, {Solanki},
  {Sch{\"u}ssler}, {Chares}, {Curdt}, {Deutsch}, {Feller}, {Germerott},
  {Grauf}, {Heerlein}, {Hirzberger}, {Kolleck}, {Meller}, {M{\"u}ller},
  {Riethm{\"u}ller}, {Tomasch}, {Kn{\"o}lker}, {Lites}, {Card}, {Elmore},
  {Fox}, {Lecinski}, {Nelson}, {Summers}, {Watt}, {Mart{\'\i}nez Pillet},
  {Bonet}, {Schmidt}, {Berkefeld}, {Title}, {Domingo}, {Gasent Blesa}, {Del
  Toro Iniesta}, {L{\'o}pez Jim{\'e}nez}, {{\'A}lvarez-Herrero},
  {Sabau-Graziati}, {Widani}, {Haberler}, {H{\"a}rtel}, {Kampf}, {Levin},
  {P{\'e}rez Grande}, {Sanz-Andr{\'e}s}, \& {Schmidt}}]{Barthol2011}
{Barthol}, P., {Gandorfer}, A., {Solanki}, S.~K., {et~al.} 2011, \solphys, 268,
  1

\bibitem[{{Bellot Rubio} \& {Orozco Su\'arez}(2019)}]{Bellot2019}
{Bellot Rubio}, L.~R. \& {Orozco Su\'arez}, D. 2019, Living Rev. Solar Phys.,
  16, 1

\bibitem[{{Bellot Rubio} {et~al.}(2006){Bellot Rubio}, Schlichenmaier, \&
  Tritschler}]{Bellot2006}
{Bellot Rubio}, L.~R., Schlichenmaier, R., \& Tritschler, A. 2006, A\&A, 453,
  1117

\bibitem[{Borrero \& {Bellot Rubio}(2002)}]{Borrero2002}
Borrero, J.~M. \& {Bellot Rubio}, L.~R. 2002, A\&A, 385, 1056

\bibitem[{Briand \& Solanki(1998)}]{Briand1998}
Briand, C. \& Solanki, S.~K. 1998, A\&A, 330, 1160

\bibitem[{Bruls {et~al.}(1992)Bruls, Rutten, \& Shchukina}]{Bruls1992}
Bruls, J. H. M.~J., Rutten, R.~J., \& Shchukina, N.~G. 1992, A\&A, 265, 237

\bibitem[{Buehler {et~al.}(2015)Buehler, Lagg, Solanki, \& {van
  Noort}}]{Buehler2015}
Buehler, D., Lagg, A., Solanki, S.~K., \& {van Noort}, M. 2015, A\&A, 576, A27

\bibitem[{Bézier(1968)}]{bezier1968}
Bézier, P. 1968, Mathematical and practical possibilities of UNISURF, renault,
  Paris

\bibitem[{Carlin(2019)}]{Carlin2019}
Carlin, E.~S. 2019, A\&A, 627, A47

\bibitem[{Cheung {et~al.}(2007)Cheung, Sch\"ussler, \&
  {Moreno-Insertis}}]{Cheung2007}
Cheung, M.~C., Sch\"ussler, M., \& {Moreno-Insertis}, F. 2007, A\&A, 461, 1163

\bibitem[{de~la Cruz~Rodr\'iguez {et~al.}(2015)de~la Cruz~Rodr\'iguez,
  L\"{o}fdahl, S\"{u}tterlin, Hillberg, \& {Rouppe van der
  Voort}}]{delaCruz2015}
de~la Cruz~Rodr\'iguez, J., L\"{o}fdahl, M.~G., S\"{u}tterlin, P., Hillberg,
  T., \& {Rouppe van der Voort}, L. 2015, A\&A, 573, A40

\bibitem[{{de la Cruz Rodr\'iguez} {et~al.}(2010){de la Cruz Rodr\'iguez},
  {Socas-Navarro}, {van Noort}, \& {Rouppe van der Voort}}]{delaCruz2010}
{de la Cruz Rodr\'iguez}, J., {Socas-Navarro}, H., {van Noort}, M., \& {Rouppe
  van der Voort}, L. 2010, Mem. Soc. Astron. Italiana, 81, 716

\bibitem[{{Dominguez Cerdeña} {et~al.}(2003){Dominguez Cerdeña}, {S\'anchez
  Almeida}, \& {Kneer}}]{Dominguez2003}
{Dominguez Cerdeña}, I., {S\'anchez Almeida}, J., \& {Kneer}, F. 2003, A\&A,
  407, 741

\bibitem[{{Dorantes-Monteagudo} {et~al.}(2022){Dorantes-Monteagudo},
  {Siu-Tapia}, {Quintero-Noda}, \& {Orozco Su\'arez}}]{Dorantes2022}
{Dorantes-Monteagudo}, A.~J., {Siu-Tapia}, A.~L., {Quintero-Noda}, C., \&
  {Orozco Su\'arez}, D. 2022, A\&A, 659, A156

\bibitem[{Farin(2002)}]{farin2002}
Farin, G.~E. 2002, Curves and Surfaces for Computer Aided Geometric Design: A
  Practical Guide, 5th edn. (San Francisco, CA: Morgan Kaufmann)

\bibitem[{Fontela {et~al.}(1993)Fontela, Avrett, \& Loeser}]{Fontela1993}
Fontela, J.~M., Avrett, E.~H., \& Loeser, R. 1993, ApJ, 406, 319

\bibitem[{{Gonz\'alez Manrique} {et~al.}(2020){Gonz\'alez Manrique}, {Quintero
  Noda}, Kuckein, {Ruiz Cobo}, \& Carlsson}]{Gonzalez2020}
{Gonz\'alez Manrique}, S.~J., {Quintero Noda}, C., Kuckein, C., {Ruiz Cobo},
  B., \& Carlsson, M. 2020, A\&A, 634, A19

\bibitem[{Jain(2010)}]{Jain2010}
Jain, A.~K. 2010, Data Clustering: A Review, Vol.~31, 264--323

\bibitem[{Jolliffe \& Cadima(2016)}]{Jolliffe2016}
Jolliffe, I.~T. \& Cadima, J. 2016, Philosophical Transactions of the Royal
  Society A: Mathematical, Physical and Engineering Sciences, 374

\bibitem[{Khomenko {et~al.}(2003)Khomenko, Collados, Solanki, Lagg, \&
  {Trujillo Bueno}}]{Khomenko2003}
Khomenko, E.~V., Collados, M., Solanki, S.~K., Lagg, A., \& {Trujillo Bueno},
  J. 2003, A\&A, 408, 1115

\bibitem[{Kuckein(2019)}]{Kuckein2019}
Kuckein, C. 2019, A\&A, 630, A139

\bibitem[{Kuckein {et~al.}(2020)Kuckein, {Gonz\'alez Manrique}, Kleint, \&
  {Asensio Ramos}}]{Kuckein2020}
Kuckein, C., {Gonz\'alez Manrique}, S.~J., Kleint, L., \& {Asensio Ramos}, A.
  2020, A\&A, 640, A71

\bibitem[{Lagg {et~al.}(2025)Lagg, Gandorfer, Solanki, \& et~al.}]{Lagg2024}
Lagg, A., Gandorfer, A., Solanki, S.~K., \& et~al. 2025, Sol. Phys.

\bibitem[{{Landi Degl'innocenti} \& {Landi
  Degl'innocenti}(1973)}]{Deglinnocenti1973}
{Landi Degl'innocenti}, E. \& {Landi Degl'innocenti}, M. 1973, Sol. Phys., 31,
  299

\bibitem[{{Landi Degl'innocenti} \& Landolfi(2004)}]{Deglinnocenti2004}
{Landi Degl'innocenti}, E. \& Landolfi, M. 2004, Polarization in spectral lines
  (United States of America: Kluwer Academic Publishers)

\bibitem[{Lites {et~al.}(1988)Lites, Skumanich, .Rees, \& Murphy}]{Lites1988}
Lites, B.~W., Skumanich, A., .Rees, D.~E., \& Murphy, G.~A. 1988, ApJ, 330, 493

\bibitem[{MacQueen(1967)}]{MacQueen1967}
MacQueen, J.~B. 1967, in Proceedings of the Fifth Berkeley Symposium on
  Mathematical Statistics and Probability, Volume 1: Statistics, University of
  California Press, 281--297

\bibitem[{{Mart\'inez Gonz\'alez} \& {Bellot Rubio}(2009)}]{Martinez2009}
{Mart\'inez Gonz\'alez}, M.~J. \& {Bellot Rubio}, L.~R. 2009, ApJ, 700, 1391

\bibitem[{{Mart\'inez Gonz\'alez} {et~al.}(2012){Mart\'inez Gonz\'alez},
  {Bellot Rubio}, {Solanki}, {Mart\'inez Pillet}, {del Toro Iniesta},
  P-Barthol, \& Schmidth}]{Martinez2012b}
{Mart\'inez Gonz\'alez}, M.~J., {Bellot Rubio}, L.~R., {Solanki}, S.~K.,
  {et~al.} 2012, APJL, 758, L40

\bibitem[{Mauas {et~al.}(1988)Mauas, Avrett, \& Loeser}]{Mauas1988}
Mauas, P.~J., Avrett, E.~H., \& Loeser, R. 1988, ApJ, 330, 1008

\bibitem[{Moe {et~al.}(2023)Moe, Pereira, Calvo, \& Leenaarts}]{Moe2023}
Moe, T.~E., Pereira, T. M.~D., Calvo, F., \& Leenaarts, J. 2023, A\&A, 675,
  A130

\bibitem[{Neckel \& Labs(1984)}]{Neckel1984}
Neckel, H. \& Labs, D. 1984, Sol. Phys., 90, 205

\bibitem[{{Quintero Noda} {et~al.}(2019){Quintero Noda}, Iijima, Katsukawa, \&
  {et al.}}]{Quintero2019}
{Quintero Noda}, C., Iijima, H., Katsukawa, Y., \& {et al.} 2019, MNRAS, 486,
  4203

\bibitem[{{Quintero Noda} {et~al.}(2022){Quintero Noda}, Schlichenmaier,
  {Bellot Rubio}, \& {et al.}}]{Quintero2022}
{Quintero Noda}, C., Schlichenmaier, R., {Bellot Rubio}, L.~R., \& {et al.}
  2022, A\&A, 666, A21

\bibitem[{{Quintero Noda} {et~al.}(2018){Quintero Noda}, Uitenbroek, Carlsson,
  {Orozco Su\'arez}, Katsukawa, {Ruiz Cobo}, Kubo, Oba, Kawabata, Hasegawa,
  Ichimoto, Anan, \& Suematsu}]{Quintero2018}
{Quintero Noda}, C., Uitenbroek, H., Carlsson, M., {et~al.} 2018, MNRAS, 481,
  5675

\bibitem[{Rimmele(2019)}]{Rimmele2019}
Rimmele, T. 2019, BAAS, 51

\bibitem[{Rimmele {et~al.}(2020)Rimmele, Warner, Keil, \& et~al.}]{Rimmele2020}
Rimmele, T.~R., Warner, M., Keil, S.~L., \& et~al. 2020, Sol. Phys., 295, 172

\bibitem[{{Ruiz Cobo} \& {del Toro Iniesta}(1992)}]{Ruiz1992}
{Ruiz Cobo}, B. \& {del Toro Iniesta}, J.~C. 1992, ApJ, 398, 375

\bibitem[{{Ruiz Cobo} {et~al.}(1996){Ruiz Cobo}, {del Toro Iniesta}, {Rodriguez
  Hidalgo}, Collados, \& {Sanchez Almeida}}]{Ruiz1996}
{Ruiz Cobo}, B., {del Toro Iniesta}, J.~C., {Rodriguez Hidalgo}, I., Collados,
  M., \& {Sanchez Almeida}, J. 1996, in ASP Conf. Ser. 109, ed. R.~Pallavicini
  \& A.~K. Dupree, Vol. 155

\bibitem[{{Ruiz Cobo} {et~al.}(2022){Ruiz Cobo}, {Quintero Noda}, Gafeira,
  Uitenbroek, {Orozco Su\'arez}, \& {P\'aez Man\'a}}]{Ruiz2022}
{Ruiz Cobo}, B., {Quintero Noda}, C., Gafeira, R., {et~al.} 2022, A\&A, 660,
  A37

\bibitem[{{Sainz Dalda} {et~al.}(2019){Sainz Dalda}, {de la Cruz Rodr\'iguez},
  {De Pontieu}, \& Go\v{s}i\'c}]{Sainz2019}
{Sainz Dalda}, A., {de la Cruz Rodr\'iguez}, J., {De Pontieu}, B., \&
  Go\v{s}i\'c, M. 2019, ApJ, 875, L18

\bibitem[{Scharmer {et~al.}(2003)Scharmer, Bjelksjo, Korhonen, Lindberg, \&
  Petterson}]{Scharmer2003}
Scharmer, G.~B., Bjelksjo, K., Korhonen, T.~K., Lindberg, B., \& Petterson, B.
  2003, SPIE Proceedings, 4853, 341

\bibitem[{{Scharmer} {et~al.}(2008){Scharmer}, {Nordlund}, \&
  {Heinemann}}]{Scharmer2008}
{Scharmer}, G.~B., {Nordlund}, A., \& {Heinemann}, T. 2008, ApJL, 677, L149

\bibitem[{Sigwarth(2001)}]{Sigwarth2001}
Sigwarth, M. 2001, ApJ, 563, 1031

\bibitem[{Sigwarth {et~al.}(1999)Sigwarth, Balasubramaniam, Kn\"olker, \&
  Schmidt}]{Sigwarth1999}
Sigwarth, M., Balasubramaniam, K.~S., Kn\"olker, M., \& Schmidt, W. 1999, A\&A,
  349, 941

\bibitem[{{Siu-Tapia} {et~al.}(2025){Siu-Tapia}, {Bellot Rubio}, \& {Orozco
  Su\'arez}}]{Siu_paperI}
{Siu-Tapia}, A.~L., {Bellot Rubio}, L.~R., \& {Orozco Su\'arez}, D. 2025, A\&A

\bibitem[{Sobotka(2003)}]{Sobotka2003}
Sobotka, M. 2003, Astron. Nachr., 4, 369

\bibitem[{Sobotka {et~al.}(2012)Sobotka, {del Moro}, Jurcak, \&
  Berrilli}]{Sobotka2012}
Sobotka, M., {del Moro}, D., Jurcak, J., \& Berrilli, F. 2012, A\&A, 537, A85

\bibitem[{Thomas \& Weiss(2004)}]{Thomas2004}
Thomas, J.~H. \& Weiss, N.~O. 2004, ARA\&A, 42, 517

\bibitem[{{van Noort} {et~al.}(2005){van Noort}, {Rouppe van der Voort}, \&
  L\"ofdahl}]{Vannoort2005}
{van Noort}, M., {Rouppe van der Voort}, L., \& L\"ofdahl, M.~G. 2005, Sol
  Phys., 228, 191

\bibitem[{Vissers \& {Rouppe van der Voort}(2012)}]{Vissers2012}
Vissers, G. \& {Rouppe van der Voort}, L. 2012, ApJ, 750, 22

\bibitem[{Viticchi\'e \& {S\'anchez Almeida}(2011)}]{Viticchie2011}
Viticchi\'e, B. \& {S\'anchez Almeida}, J. 2011, A\&A, 530, A14

\bibitem[{Vukadinovic {et~al.}(2022)Vukadinovic, Milic, \&
  Atanackovic}]{Vukadinovic2022}
Vukadinovic, D., Milic, I., \& Atanackovic, O. 2022, A\&A, 664, A182

\bibitem[{Wold {et~al.}(1987)Wold, Esbensen, \& Geladi}]{Wold1987}
Wold, S., Esbensen, K., \& Geladi, P. 1987, Chemometrics and Intelligent
  Laboratory Systems, 2, 37

\end{thebibliography}

%

\begin{appendix}



\onecolumn
\section{Additional material}\label{Ap:A}

\begin{table}[h!]
\centering 
\caption{
Inversion configuration for the $V_R$ representative profiles.}        
\label{tab1}      
\centering                                      
\begin{tabular}{c c c c c c c c c c c}          
\hline\hline                        
$F$ & $V_R$ & $L$ & $W$ & $C$ & $T$ & $v_{mic}$ & $v_{LOS}$ & $B$ & $\gamma$ & $\phi$ \\
 &  &  & $I,Q,U,V$ &  &  & &  &  &  &  \\

\hline                                   
\hline                                             

 \multirow{2}{*}{A} & 0 & BS & 10,1,1,10 & 4 & 5 & 3 & 2 & 2 & 1 & 1 \\ 
 & 1 & BS & 10,1,1,10 & 5 & 5 & 5 & 3 & 3 & 2 & 2 \\ 
 \hline
 \multirow{2}{*}{B} & 2 & PP & 10,1,1,10 & 4 & 5 & 5 & 3 & 2 & 2 & 2 \\ 
 & 3 & PP & 10,1,1,10 & 7 & 9 & 5 & 5 & 2 & 2 & 2 \\
 \hline
\multirow{3}{*}{C}  & 4 & PC & 10,1,1,10 & 12 & 6 & 6 & 3 & 5 & 3 & 2 \\
  & 5 & PC & 10,1,1,10 & 8 & 9 & 5 & 5 & 5 & 2 & 2 \\ 
 & 6 & PC & 1,1,1,1 & 8 & 5 & 3 & 6 & 6 & 2 & 2 \\ 
 \hline
  \multirow{4}{*}{D} &  7 & BS & 10,1,1,10 & 5 & 3 & 2 & 2 & 2 & 2 & 2 \\ 
    &  &  &  &  & 3 & 2 & 2 & 2 & 2 & 2 \\ 
 & 8 & BS & 10,1,1,10 & 5 & 5 & 2 & 2 & 5 & 2 & 2 \\ 
    &  &  &  &  & 5 & 2 & 5 & 2 & 2 & 2 \\ 
\hline                                   
\hline     
\end{tabular}
\tablefoot{From left to right, the columns indicate the Family $F$, the representative profile, the location $L$ of the representative profile (QS$\equiv$Quiet Sun, BS$\equiv$Bright Structure, 
PP$\equiv$Pore Periphery, PC$\equiv$Pore Center; see also Fig.\ \ref{fig:4bef}), the weight given to each of the Stokes parameters $W$, the number of cycles $C$, and the maximum number of nodes used for each of the physical parameters in the model atmosphere. The inversions of $V_R$ 7 and 8 consider two atmospheric components, and are described in two different rows.}
\end{table}

\begin{table}[h!]
\centering 
\caption{
Inversion configuration for  the $I_R$ representative profiles.}        
\label{tab2}      
\centering                                      
\begin{tabular}{c c c c c c c c c c c}          
\hline\hline                        
$F$ & $I_R$ & $L$ & $W$ & $C$ & $T$ & $v_{mic}$ & $v_{LOS}$ & $B$ & $\gamma$ & $\phi$ \\
 &  &  & $I,Q,U,V$ &  &  & &  &  &  &  \\

\hline                                   
\hline                                             
\multirow{3}{*}{A} & 0 & QS & 1,1,1,1 & 3 & 5 & 3 & 2 & 1 & 1 & 1 \\ 
 & 1 & QS & 1,1,1,1 & 3 & 5 & 3 & 3 & 2 & 1 & 1 \\ 
 & 2 & QS & 1,1,1,1 & 3 & 5 & 3 & 2 & 1 & 1 & 1 \\ 
 \hline
 \multirow{1}{*}{B} & 4 & BS & 10,1,1,10 & 6 & 6 & 6 & 3 & 3 & 2 & 2 \\ 
 \hline
 \multirow{1}{*}{C} & 5 & PP & 1,1,1,1 & 7 & 11 & 9 & 3 & 2 & 2 & 2 \\ 
  \hline
 \multirow{2}{*}{D} & 7 & PC & 10,1,1,10 & 8 & 9 & 3 & 3 & 3 & 3 & 2 \\ 
 & 8 & PC & 1,1,1,1 & 8 & 11 & 11 & 3 & 9 & 9 & 5 \\ 
   \hline
  \multirow{10}{*}{E}  & 10 & BS & 10,1,1,10 & 6 & 5 & 2 & 2 & 2 & 2 & 2 \\ 
 &  &  &  &  & 5 & 2 & 2 & 2 & 2 & 2 \\ 
 & 11 & BS & 10,1,1,10 & 6 & 5 & 5 & 5 & 3 & 2 & 2 \\ 
 &  &  &  &  & 5 & 5 & 5 & 3 & 2 & 2 \\ 
 & 12 & QS & 1,1,1,1 & 10 & 5 & 6 & 6 & 3 & 2 & 2 \\ 
 &  &  &  &  & 5 & 6 & 6 & 3 & 2 & 2 \\ 
  & 13 & QS & 1,1,1,1 & 8 & 6 & 6 & 9 & 5 & 3 & 2 \\ 
 &  &  &  &  & 6 & 6 & 9 & 5 & 3 & 2 \\ 
  & 14 & QS & 10,1,1,10 & 8 & 5 & 3 & 5 & 5 & 2 & 2 \\ 
 &  &  &  &  & 5 & 3 & 5 & 5 & 2 & 2 \\ 
  & 15 & QS & 1,1,1,1 & 8 & 6 & 6 & 5 & 3 & 3 & 3 \\ 
 &  &  &  &  & 6 & 6 & 5 & 3 & 3 & 3 \\ 
  
\hline                                   
\hline     
\end{tabular}
\tablefoot{The format is the same as in Table \ref{tab1}.  
The inversions of 
$I_R$ 10 to $I_R$ 15
consider two atmospheric components, and are described in two different rows.}
\end{table}

 \begin{figure*}[h!]
   \centering
 
       \includegraphics[width=\hsize]{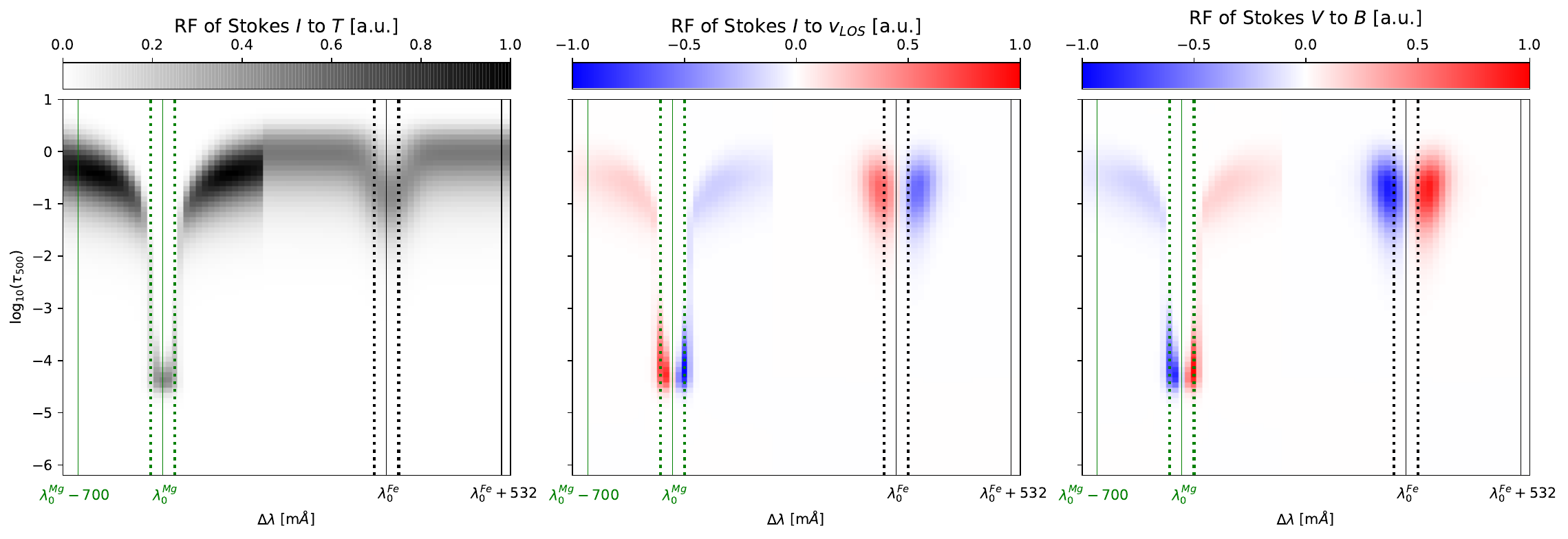} 
        \includegraphics[width=\hsize]{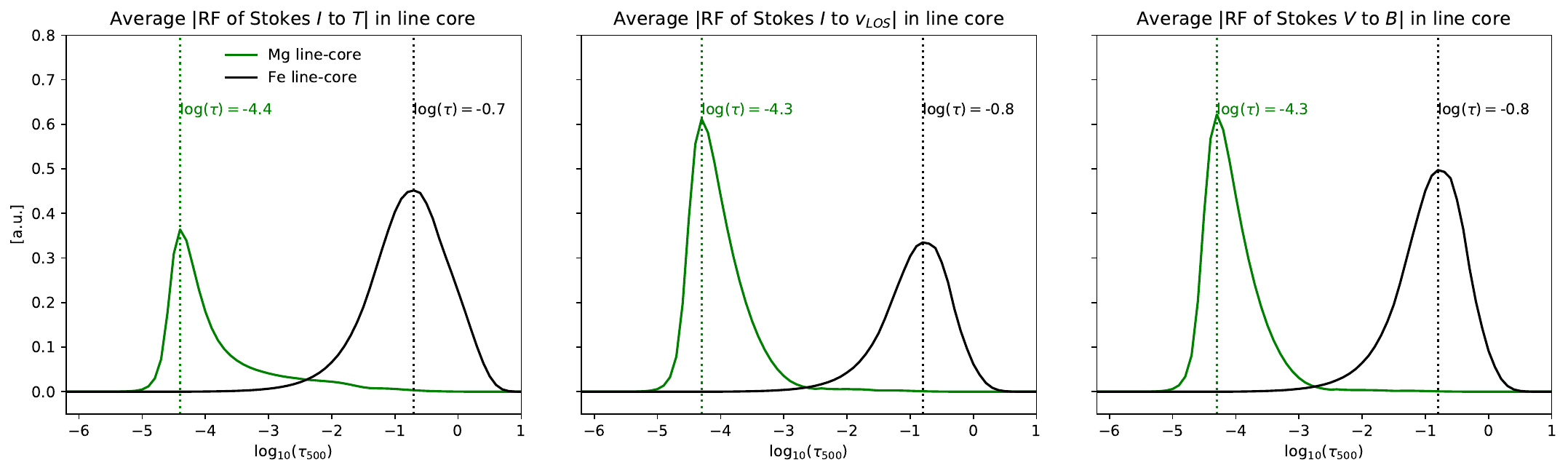} 
       
      \caption{
  Top: Response functions to temperature (left), LOS velocity (middle), and magnetic field strength (right) of the Mg I 5173 \AA\ and Fe I 6173 \AA\ lines computed for $V_R$ 0 with DeSIRe, using the model resulting from the NLTE inversions displayed in Fig.\ \ref{fig:15}. All the cases are normalized to the maximum of the RF of Stokes $I$ to changes in the temperature. Green dotted lines delimit a spectral window of $\pm 100$ m\AA\ around the  core of the Mg line, while black dotted lines enclose the core of the Fe line in the spectral range $\pm 56$ m{\AA}. Bottom: Average of the unsigned RFs within the spectral windows containing the core of the Mg line (green) and the core of the Fe line (black) lines as defined in the upper plots. The vertical dashed lines are placed at the optical depths of maximum average response in each case. 
              }
         \label{fig:15rfs}
   \end{figure*}

\end{appendix}

\end{document}